\documentclass[prd,superscriptaddress,nofootinbib,showpacs,preprint]{revtex4}
\usepackage{graphicx}
\usepackage{dcolumn}
\usepackage{bm}
\usepackage{amssymb}
\usepackage{mathrsfs}
\usepackage{amsmath}
\usepackage{epsfig}
\usepackage[dvips]{color}
\usepackage{hhline}
\usepackage[dvipdfm,colorlinks,citecolor=blue]{hyperref}
\usepackage[normalem]{ulem}

\def\lapp{\mathrel{\rlap{\raise.5ex\hbox{$<$}}
                    {\lower.5ex\hbox{$\sim$}}}}
\def\gapp{\mathrel{\rlap{\raise.5ex\hbox{$>$}}
                    {\lower.5ex\hbox{$\sim$}}}}

\usepackage{subfigure}
\usepackage{times}
\newcommand{\bmt}{\begin{pmatrix}}
\newcommand{\emt}{\end{pmatrix}}

\usepackage{multirow}
\newcommand{\ba}{\begin{array}{c}}
\newcommand{\ea}{\end{array}}
\newcommand{\be}{\begin{equation}}
\newcommand{\ee}{\end{equation}}
\newcommand{\bea}{\begin{eqnarray}}
\newcommand{\eea}{\end{eqnarray}}

\newcommand{\bi}{\begin{itemize}}
\newcommand{\ei}{\end{itemize}}
\begin{document}
\title{Sub-dominant Type II Seesaw as an Origin of Non-zero $\theta_{13}$ in SO(10) model 
with TeV scale $Z^\prime$ Gauge Boson}
\author{Debasish Borah}
\email{dborah@tezu.ernet.in}
\affiliation{Department of Physics, Tezpur University, Tezpur - 784028, India}
\author{Sudhanwa Patra}
\email{sudha.astro@gmail.com}
\affiliation{Center of Excellence in Theoretical and Mathematical Sciences, 
Siksha 'O' Anusandhan University, Bhubaneswar-751030, India}
\author{Prativa Pritimita}
\email{pratibha.pritimita@gmail.com}
\affiliation{Center of Excellence in Theoretical and Mathematical Sciences, 
Siksha 'O' Anusandhan University, Bhubaneswar-751030, India}
\begin{abstract}
We discuss a class of left-right symmetric models where the light neutrino masses originate 
dominantly from type I seesaw mechanism along with a sub-dominant type II seesaw contribution. 
The dominant type I seesaw gives rise to tri-bimaximal type neutrino mixing whereas sub-dominant 
type II seesaw acts as a small perturbation giving rise to non-zero $\theta_{13}$ in our model 
which also has TeV scale right-handed neutrinos and $Z^\prime$ gauge boson thereby making the 
model verifiable at current accelerator experiments. Sub-dominant type II and dominant type I 
seesaw can be naturally accommodated by allowing spontaneous breaking of D-parity and $SU(2)_R$ 
gauge symmetry at high scale and allowing TeV scale breaking of $U(1)_{R} \times U(1)_{B-L}$ 
into $U(1)_Y$. We also embed the left-right model in a non-supersymmetric $SO(10)$ grand unified 
theory (GUT) with verifiable TeV scale $Z^\prime$ gauge boson. Drawing it to an end, we scrutinize 
in detail the evaluation of one-loop renormalization group evolution for relevant gauge couplings 
and estimation of the proton life time which can be accessible to the foreseeable experiments. 
And in the aftermost part we make an estimation of branching ratio for lepton flavor violating 
process $\mu \rightarrow e + \gamma $ as a function of type II seesaw strength due to doubly 
charged component of the right handed Higgs triplet with mass at the TeV scale, which can be 
accessible at ongoing experiments.
\end{abstract}
\maketitle
\section{Introduction}
The fact that the most successful phenomenological theory, the standard model (SM) of particle physics, 
suffers from the inability to address several observed phenomena as well as theoretical questions, has 
always been a source of excitement for particle physicists. The tiny but non-zero neutrino masses that 
have been confirmed by the phenomenon of neutrino oscillations detected in solar, atmospheric and reactor 
experiments \cite{neutosc} is certainly one such phenomena which the SM fails to address. These observations, 
among others have intensified the urge to ponder beyond the SM which has led to several well motivated 
beyond SM frameworks. The canonical seesaw mechanism (commonly referred to as type-I seesaw \cite{ti}), 
being the most elegant mechanism for generating small neutrino masses relies on the existence of 
right-handed (RH) neutrinos. Fundamentally speaking, the RH neutrinos are singlets under the SM gauge 
symmetry and hence can have arbitrary (can be very large) Majorana masses leading to light neutrino 
masses as $m^I_\nu \simeq y_\nu v^2/M_N$, where $y_\nu$ is the Dirac Yukawa coupling, $v=246$ GeV 
is the vacuum expectation value (VEV) of the SM Higgs and $M_N$ is the RH Majorana neutrino mass. 
In order to be compatible with the neutrino oscillation data, i.e, $|m_\nu| \simeq \sqrt{\Delta 
m^2_{\rm atm}}=0.0495$ eV, one requires $M_N\simeq 10^{14-15}$ GeV taking the Yukawa couplings 
in their natural values i.e, $\mathcal{O}(1)$.

In addition to the canonical seesaw, other seesaw mechanisms have been worked upon as well 
to explain the tiny masses of the active neutrinos. Picking the type-II seesaw mechanism 
\cite{tii} from them which requires the existence of $SU(2)_L$ triplet Higgs fields in addition 
to the minimal SM particle content, the neutrino mass gets an extra contribution given by 
$m^{II}_\nu \simeq f\, v_L$, where $v_L$ is the VEV of the neutral component of the triplet 
and $f$ is the corresponding Yukawa coupling. Minimizing the scalar potential of such a model, 
the VEV of the Higgs triplet is found to be $v_L=\mu v^2/M^2_{\Delta}$, 
where $M_{\Delta}$ is the mass of the Higgs triplet and $\mu$ defines the mixing between 
SM Higgs and triplet. An obvious setting would be $f\simeq \mathcal{O}(1)$ and $\mu \sim M_\Delta 
\simeq 10^{14-15}$ GeV in order to explain sub-eV scale of light neutrino masses. Although these 
seesaw mechanisms look promising while explaining neutrino oscillation data, they lack the direct 
experimental testability in the ongoing experiments like Large Hadron Collider (LHC) or any future 
experiment like International Linear Collider.

Retreating not here, the particle phenomenology community has explored  beyond standard model 
physics operative at few TeV scale, the results being repetitive attempts to corroborate neutrino 
masses and mixing. One such highly motivated and one of the most widely discussed beyond standard 
model framework is the left-right symmetric model (LRSM) \cite{LR} which not only gives a clear 
description of the origin of parity violation at electroweak scale but also leads a way to the 
generation of neutrino masses naturally. A thorough study of these models strengthens us with 
the knowledge that in conventional left-right symmetric model (LRSM), the light neutrino masses 
arise from two sources: the type-I \cite{ti} plus type-II \cite{tii} seesaw mechanisms 
where the parity and $SU(2)_R$ gauge symmetry are spontaneously broken at the same scale.

Earlier explorations of the field imply that a deep relation between high energy collider physics 
and low energy phenomena like neutrino-less double beta decay as well as other lepton flavor violating 
processes is enrooted by minimal left-right symmetric model (LRSM) valid at TeV scale \cite{tello, patra-jhep}. 
If it happens that the parity and $SU(2)_R$ break at the same scale, then according to the seesaw 
relation $v_L v_R = \gamma v^2$ (with $\gamma$ being a dimensionless parameter) the microscopic value 
of $v_L$ as required for type-II seesaw depends on large value of $v_R$, which further implies $v_R\approx 
\left(10^{13} \sim 10^{14}\right)$ GeV making itself incapable of direct detection in near future. 
In a contrast way, if the right-handed scale is assigned with more moderate values, say in the range 
of few TeV, one can expect to have observable consequences at experiments in the near future. 
More willingly, if we assume both parity and $SU(2)_R$ to be broken at TeV scale i.e, $v_R \simeq$ 
TeV that is the scale of RH heavy neutrino mass, then we strictly need to calibrate the Higgs couplings 
up to the order of $\gamma \leq {\cal O}(10^{-10})$ in order to fit neutrino data from the seesaw 
relation. To refine this, studies have been done on left-right symmetric models to come upon with 
spontaneous D-parity breaking \cite{Dparity11,Dparity22,utpal-desai,utpal-sahu,patra-debi} where parity 
gets broken much earlier than $SU(2)_R$ gauge symmetry. In this work, we shall be discussing such a class 
of left-right symmetric models in which the spontaneous breaking of D-parity occurs at reasonably high 
scale along with $SU(2)_R \times U(1)_{B-L}$ gauge symmetry breaking down to $U(1)_R \times U(1)_{B-L}$. 
We then check numerically whether $U(1)_R \times U(1)_{B-L}$ breaking occurs at TeV scale (provided parity 
breaks at much higher scale) and tiny neutrino masses can be obtained without too much fine tuning. In this 
class of models, the TeV scale breaking of $U(1)_R \times U(1)_{B-L}$ results in the TeV scale masses 
of the right-handed neutrinos as well as $Z^{\prime}$ boson while D-parity breaks at a high energy scale 
($\simeq 10^{9-11}$ GeV). As will be discussed later, this allows the possibility of dominant type I seesaw 
contribution to neutrino mass whereas type II seesaw contribution can naturally remain sub-dominant. We use 
such a sub-dominant type II seesaw contribution as the origin of non-zero $\theta_{13}$, the reactor mixing 
angle. It should be noted that, most of the earlier attempts to explain the non-zero $\theta_{13}$ incorporate 
different corrections to the $\mu-\tau$ symmetric tri-bimaximal (TBM) neutrino mass matrix which can naturally 
originate in generic flavor symmetry models like $A_4$. Motivated by this, we consider the dominant type I 
seesaw contribution giving rise to TBM type neutrino mass matrix whereas the sub-dominant type II term giving 
rise to non-zero $\theta_{13}$. We also constrain the D-parity breaking scale from the demand of generating 
the experimentally allowed range of $\theta_{13}$. Apart from this, we also investigate whether such a choice 
of intermediate symmetry breaking scales allows the possibility to unify all the gauge couplings while being 
embedded in a non-supersymmetric $SO(10)$ grand unified theory. 

With all these motivations, we present a $SO(10)$ model with a novel chain of symmetry breaking having 
left-right symmetry as an intermediate step giving neutrino masses through type-I plus type-II seesaw 
mechanisms, unification of three fundamental forces, prediction of proton life time accessible to the 
ongoing search experiments and most importantly, a low mass $Z^\prime$ gauge boson which can be probed 
at LHC. While preparing this manuscript, an interesting work appeared online \cite{parida1312} with 
similar symmetry breaking chains and scales as the one we are discussing here. However, the neutrino 
mass phenomenology in that work is completely different from the one we pursue here. The plan of the paper 
can be sketched as follows. In section \ref{lrsmreview} we briefly discuss the left-right symmetric models, 
elucidating the spontaneous breaking of D-parity. In section \ref{numass} we discuss neutrino masses and 
mixing via dominant type-I seesaw giving rise to TBM type neutrino mixing and sub-dominate type-II seesaw 
giving rise to deviations from TBM mixing and hence non-zero $\theta_{13}$. In Sections \ref{embedso10} 
and \ref{unifso10}, we give a possible path for embedding the present left-right symmetric models in the 
non-SUSY $SO(10)$ GUT with its symmetry breaking pattern and one-loop gauge coupling unification. 
In Section \ref{pdecay}, the proton lifetime is estimated using the value gauge coupling at GUT scale. 
In section \ref{LFV}, we estimate the branching ratio for lepton flavor violating decay $\mu \rightarrow 
e+\gamma$ as a function of type II seesaw strength and finally conclude in section \ref{conclude}.

\section{Left-right symmetric model with spontaneous D-parity breaking}
\label{lrsmreview}
In left-right symmetric models with spontaneous D-parity breaking, the discrete symmetry called D-parity 
gets broken earlier compared to the $SU(2)_R$ gauge symmetry. Here the gauge group can be written effectively 
as $SU(2)_L \times SU(2)_R \times U(1)_{B-L} \times SU(3)_C \times D \,(\mathcal{G}_{2213D})$, where $D$ 
is the discrete left-right symmetry or D-parity. In matter sector, the left and right handed fermions 
are doublets under $SU(2)_L$ and $SU(2)_R$ gauge groups, respectively. The transformation of quarks 
and leptons under the left-right symmetric group can be summarized as 
\begin{eqnarray}
Q_{L}=\begin{pmatrix}u_{L}\\
d_{L}\end{pmatrix}\equiv[2,1,{\frac{1}{3}}, 3] & , & Q_{R}=\begin{pmatrix}u_{R}\\
d_{R}\end{pmatrix}\equiv[1,2,{\frac{1}{3}},3]\,,\nonumber \\
\ell_{L}=\begin{pmatrix}\nu_{L}\\
e_{L}\end{pmatrix}\equiv[2,1,{-1}, 1] & , & \ell_{R}=\begin{pmatrix}N_{R}\\
e_{R}\end{pmatrix}\equiv[1,2,-1,1] \nonumber
\end{eqnarray}

Notably the difference between Lorentz parity and D-parity is that Lorentz parity acts on the Lorentz 
group and interchanges left-handed fermions with the right-handed ones but the bosonic fields remain 
the same whereas $D$-parity acts on the gauge groups $SU(2)_L \times SU(2)_R$ interchanging the $SU(2)_L$ 
Higgs fields with the $SU(2)_R$ Higgs fields in addition to the interchange of fermions. The spontaneous 
breaking of D-parity creates an asymmetry between left and right handed Higgs fields making the coupling 
constants of $SU(2)_R$ and $SU(2)_L$ evolve separately under the renormalization group. 

The Higgs sector of the left-right model with spontaneous D-parity breaking mechanism 
consists of a $SU(2)$ singlet scalar field $\sigma$ which is odd under discrete D-parity, 
two $SU(2)_L$ triplets $\Delta_L, \Delta_R$ and a bidoublet $\Phi$ which contains two copies of $SM$ 
Higgs transforming under the LR gauge group $\mathcal{G}_{2213}=SU(2)_L \times SU(2)_{R} \times 
U(1)_{B-L} \times SU(3)_C$ as
\begin{eqnarray}
& &\Delta_{L} = 
(3,1,-2,1)\,, 
\Delta_{R} = 
(1,3,-2,1)\, , \nonumber \\
& &\Phi =
(2,2,0,1) \,, \sigma=(1,1,0,1)\,. \nonumber 
\end{eqnarray}

By assigning a non-zero VEV to D-parity odd singlet $\langle \sigma \rangle \simeq M_P$, the left-right symmetry 
is spontaneously broken but the gauge symmetry $\mathcal{G}_{2213}$ remains unbroken resulting in 
\begin{eqnarray}
&&M^2_{\Delta_R}=M^2_\Delta - \lambda_{\Delta} \langle \sigma \rangle M\, , \nonumber \\
&&M^2_{\Delta_L}=M^2_\Delta + \lambda_{\Delta} \langle \sigma \rangle M\, ,
\end{eqnarray}
where $M_\Delta$ is the mass term for triplets i.e, $M^2_{\Delta} \mbox{Tr}\left(\Delta^\dagger_L \Delta_L+ 
\Delta^\dagger_R \Delta_R\right)$, 
and $\lambda_\Delta$ is the trilinear coupling in the term $M \sigma \mbox{Tr}\left(\Delta^\dagger_L \Delta_L- 
\Delta^\dagger_R \Delta_R\right)$. In this scenario $M_{\Delta}, M, \langle \sigma \rangle$ all are of order of $M_P$ which is the scale of 
D-parity breaking thereby resulting TeV scale masses for right-handed Higgs triplets and D-parity breaking scale 
for their left-handed counterparts by suitable adjsutment of trilinear coupling $\lambda_{\Delta}$. 
In order to have $W_R$ and $Z_R$ mass predictions at nearly the same scale 
along with the generation of Majorana neutrino masses, it is customary to break $SU(2)_{R} \times U(1)_{B-L} 
\rightarrow U(1)_Y$ in a single step by the VEV of the right handed triplet $\langle\Delta^0_R\rangle\sim v_R$.

Instead of pursuing the aforementioned left-right symmetric model with D-parity breaking mechanism, we consider a more appealing 
phenomenological scenario: 
\begin{equation}
G_{2213D} \stackrel{M_P}{\longrightarrow}  G_{2113} \stackrel{M^0_R}{\longrightarrow} 
\mathcal{G}_{213}( \mbox{SM})
\mathop{\longrightarrow}^{\Phi} \mathcal{G}_{13}                           
\end{equation}
with $M_{W_R} >> M_{Z_R}$ via two step breaking of the left-right symmetric gauge 
theory to the SM. The Higgs sector of the present model with spontaneous D-parity breaking mechanism 
consists of two $SU(2)_L$ triplets $\Delta_L$ and $\Omega_L$, two $SU(2)_R$ triplets $\Delta_R$, $\Omega_R$ 
and a bidoublet $\Phi$ which contains two copies of $SM$ Higgs transforming under the LR gauge groups 
is shown in Table.\,\ref{table:Higgs:LR}. 
\begin{table}[h!]
\begin{center}
\begin{tabular}{|c|c|c|}
\hline \hline
Higgs Fields & Under $\mathcal{G}_{2213}$& Under $\mathcal{G}_{2113}$\\
 & $\left(2_L,2_R,1_{B-L},3_C \right)$ 
& $\left(2_L,1_R,1_{B-L},3_C \right)$   \\
\hline
$\Omega_R$ 
& $\left[1,3,0,1 \right]$ 
& $\left[1,1,0,1 \right]$ 
\\
\hline
$\Omega_L$ 
& $\left[3,1,0,1 \right]$ 
& $\left[3,0,0,1 \right]$ 
\\
\hline
$\Delta_R$ 
& $\left[1,3,-2,1 \right]$ 
& $\left[1,1,-2,1 \right]$ 
\\
\hline
$\Delta_L$ 
& $\left[3,1,-2,1 \right]$ 
& $\left[3,0,-2,1 \right]$ 
\\
\hline
$\Phi$ 
& $\left[2,2,0,1 \right]$ 
& $\left[2, \pm 1/2, 0,1 \right]$ 
\\
\hline \hline
\end{tabular}
\end{center}
\caption{The Higgs fields transform under relevant gauge group as $\mathcal{G}_{2213}={\small SU(2)_L \times 
SU(2)_{R} \times U(1)_{B-L} \times SU(3)_C}$ and $\mathcal{G}_{2113}={\small SU(2)_L \times 
U(1)_{R} \times U(1)_{B-L} \times SU(3)_C}$. We have chosen those fields in the third column 
under $\mathcal{G}_{2113}$ which acquire 
a non-zero vacuum expectation value and in particular, the $U(1)_R$ values corresponds to the z-components of 
Isospin i.e, $T_{3R}$ of $SU(2)_R$ satisfying $Q=T_{3L}+T_{3R}+(B-L)/2$ valid both for $\mathcal{G}_{2113}$ as well 
as $\mathcal{G}_{2213}$ gauge groups.}
\label{table:Higgs:LR}
\end{table}

\newpage
The first step of symmetry breaking i.e, $SU(2)_L \times SU(2)_{R} \times U(1)_{B-L} \times SU(3)_C \times D 
\to SU(2)_L \times U(1)_{R} \times U(1)_{B-L} \times SU(3)_C$ occurs at $W_R$ boson mass scale which is implemented 
through the VEV of the heavier triplet carrying $B-L=0$ i.e, $\langle \Omega^0_R(1,3,0,1) \rangle$ around D-parity 
breaking scale $M_P$. The second step of breaking $SU(2)_L \times U(1)_{R} \times U(1)_{B-L} \times SU(3)_C \to 
\mathcal{G}_{\rm SM}$ occurs at $Z_R$ 
mass scale and is carried out by $\langle\Delta^0_R(1,1,-2,1)\rangle\sim v_R$ around $M^0_R\simeq (3-5)$ TeV. 
This unique scenario gives us the knowledge 
that $W_R$ scale completely decouples from $Z_R$ scale and hence, the LHC signatures of these gauge bosons and corresponding bounds 
on their mass scales should be revived again. The right handed neutral gauge boson $Z_R$ gets mass around few TeV staying very close to the experimental lower bound $M_{Z^{\prime}}\ge 1.162$ TeV allowing its visibility at high energy accelerators in near future.

Apart from the right handed triplets whose VEV give masses to the right handed gauge bosons, the left handed triplets can also acquire non-zero VEV due to several scalar mixing terms in the Lagrangian. The analytic expression for VEV of the neutral component of $\Delta_L$ can be expressed as 
\begin{equation}
v_L \approx \frac{\beta v^2v_R}{2M\, M_P}\,,
\label{vev_value} 
\end{equation}
where we have used $v=246$ GeV and $\beta$ is a coupling constant of $\mathcal{O}(1)$. Noticeably 
in the above eq.(\ref{vev_value}), the smallness of the VEV of $\Delta_L$ is decided by the parity 
breaking scale and not by the $SU(2)_R$ breaking scale thereby putting no constraints on $v_R$ from 
the type-II seesaw point of view. Therefore, the type-II seesaw relation is modified for left-right 
models accompanied by spontaneous D-parity breaking scenario instead of its usual expression valid 
for conventional left-right symmetric model.  As a result, the type-I \cite{ti} seesaw term decouples 
completely from D-parity breaking scale and become sensitive to the  $U(1)_R \times U(1)_{B-L}$ breaking 
scale $M^0_R$ while the type-II \cite{tii} seesaw contribution becomes sensitive to the D-parity breaking scale. 
In the following section we shall briefly discuss how a particular value of D-parity breaking scale 
$M_P=10^{9}-10^{10}$ GeV leads to sub-dominant type-II seesaw giving rise to correct deviations from 
TBM neutrino mixing in order to generate non-zero $\theta_{13}$. As we show later, the D-parity breaking scale $M_P \sim M$ is constrained to be greater than around $3 \times 10^9$ GeV. Hence, for $v_R \sim 1$ TeV and order one dimensionless couplings, the type II contribution comes out to be $0.001$ eV or less. The leading order TBM type neutrino 
mass matrix can originate from usual type I seesaw term due to the TeV scale right handed neutrinos 
originating from the TeV scale breaking of $U(1)_{B-L}$. 
\section{Neutrino Mass}
\label{numass}
The renormalizable invariant Yukawa Lagrangian that gives rise to the $G_{2113}$ invariant interactions, 
near the TeV scale for the model considered in our present analysis, is
\begin{eqnarray}
\mathcal{L}_{\rm Yuk}&= & Y_{\ell} \overline{\ell}_L\, N_R\, \Phi + f_R\, N^c_R\, N_R \Delta_R  
+ f_L\, \nu^c_L\, \nu_L \Delta_L
                       +\,  \text{h.c.}\, ,\nonumber 
\end{eqnarray}
resulting in $6\times 6$ neutral fermion mass matrix after electroweak symmetry breaking
\begin{equation}
\mathcal{M}_\nu= \left( \begin{array}{cc}
              M_{LL} & M_{LR}   \\
              M^T_{LR} & M_{RR}
                      \end{array} \right) \, ,
\label{eqn:numatrix}       
\end{equation}
One should note here that all the mass scales used in above mass matrix $\mathcal{M}_\nu$ have 
their dynamical interpretations in this model like $M_{RR}=f_R\,v_R$, $M_{LL}=f_L\,v_L$, and 
$M_{LR}=y_\nu\,v$ in contrast to the SM where two of them $M_{LL}$, $M_{RR}$ have no dynamical 
origins. The resulting light neutrino mass can be written as a seesaw formula given by
\begin{equation}
m_{LL}=m_{LL}^{II} + m_{LL}^I
\label{type2a}
\end{equation}
 where the usual  type I  seesaw formula  is given by the expression,
\begin{equation}
m_{LL}^I=-M_{LR}M_{RR}^{-1}M_{LR}^{T}.
\end{equation}
Here $M_{LR}$ is the Dirac neutrino mass matrix. Thus, for type I seesaw dominance with TeV scale $U(1)_{B-L}$ breaking $v_R \sim 1$ TeV, the Dirac Yukawa copulings should be fine tuned to $y_{\nu} \sim 10^{-5}$ for $f_R \sim 1$. The type II seesaw term ($m_{LL}^{II} = f_L v_L$) however, is directly proportional to the Majorana Yukawa couplings $f_L$ which have to be large in order to have sizeable contribution to neutrino masses.

\begin{table}[h!]
\begin{tabular}{ccc}\\
        \hline 
parameter   & best-fit            &  3$\sigma$  \\
        \hline
$\Delta m^2_{\rm {21}} [10^{-5} \mbox{eV}^2]$ & 7.50 & 7.00-8.09 \\
$|\Delta m^2_{\rm {31}}(\mbox{NH})| [10^{-3} \mbox{eV}^2]$ & 2.473  & 2.27-2.69 \\
$|\Delta m^2_{\rm {23}}(\mbox{IH})| [10^{-3} \mbox{eV}^2]$ & 2.42  & 2.24-2.65 \\
$\sin^2\theta_{12}$  & 0.306 & 0.27-0.34     \\
$\sin^2\theta_{23}$  & 0.42  & 0.34-0.67     \\
$\sin^2\theta_{13}$  & 0.021 & 0.016-0.030   \\
        \hline
\end{tabular}
\caption{The global fit values for the mass squared differences and mixing angles taken 
from \cite{schwetz12}}
\label{table-osc}
\end{table} 

The induced VEV for the left handed triplet $v_{L}$ can be shown for generic 
LRSM to be $$v_{L}=\gamma \frac{M^{2}_{W}}{v_{R}}$$. This expression for type II seesaw term 
is valid for those class of minimal models where D-parity and $SU(2)_R\times U(1)_{B-L}$ gauge 
symmetry get broken spontaneously at the same energy scale. However, as discussed in the previous 
section, it is possible to break D-parity and $SU(2)_R\times U(1)_{B-L}$ gauge symmetry at two 
different stages. In the left right symmetric models discussed in the previous sections, 
D-parity and $SU(2)_R$ gauge symmetry get broken down to $U(1)_R$ at a very high scale whereas 
$U(1)_R \times U(1)_{B-L}$ gets broken down to $U(1)_Y$ of standard model at TeV scale. 
The VEV of the left handed triplet is given by equation (\ref{vev_value}) in such a case.

Before doing a numerical analysis of neutrino mass and mixing in our model, we note that prior 
to the discovery of non-zero $\theta_{13}$, the neutrino oscillation data were compatible with 
the well motivated TBM form of the neutrino mixing matrix discussed extensively in the literature 
\cite{Harrison}. However, since the latest data (last five references in \cite{neutosc}) have 
ruled out $\text{sin}^2\theta_{13}=0$, one needs to go beyond the TBM framework to incorporate 
non-zero $\theta_{13}$. Since the experimental value of $\theta_{13}$ is much smaller than atmospheric 
and solar neutrino mixing angles, TBM type mixing can still be a valid approximation and the non-zero 
$\theta_{13}$ can be accounted for by incorporating small perturbations to TBM mixing coming from 
different mechanisms like charged lepton mass diagonalization, for example. There have already been 
a great deal of activities in this context \cite{nzt13, nzt13GA} which can successfully explain 
the latest data within the framework of several interesting models.

Since non-zero $\theta_{13}$ can be very naturally explained by incorporating corrections to TBM 
mixing and our model naturally provides such small correction in the form of type II seesaw term, 
we find it interesting to explore the possibility of TBM type mixing coming from type I seesaw term 
and the origin of non-zero $\theta_{13}$ through the type II seesaw term. Similar attempts to study 
the deviations from TBM mixing by using the interplay of two different seesaw mechanisms were done in 
\cite{devtbmt2, dborah7-13}. Our analysis here differs from these in the sense that we implement our 
model within a grand unified theory where the strength of seesaw terms can be naturally explained from 
gauge coupling unification point of view. We also extend our earlier discussion \cite{dborah7-13} 
to include two different cases: one where the light neutrinos are almost degenerate, and the other 
in which there exists a moderate hierarchy between them, both obeying the cosmological upper limit 
on the sum of absolute neutrino masses.

Type I seesaw giving rise to $\mu-\tau$ symmetric TBM mixing pattern for neutrinos have been discussed 
extensively in the literature. The neutrino mass matrix in these scenarios can be written in a parametric 
form as
\begin{equation}
m_{LL}=\left(\begin{array}{ccc}
x& y&y\\
y& x+z & y-z \\
y & y-z & x+z
\end{array}\right)
\label{matrix1}
\end{equation}
which is clearly $\mu-\tau$ symmetric with eigenvalues $m_1 = x-y, \; m_2 = x+2y, \; m_3 = x-y+2z$. 
It predicts the mixing angles as $\theta_{12} \simeq 35.3^o, \; \theta_{23} = 45^o$ and $\theta_{13} = 0$. 
Although the prediction for first two mixing angles are still allowed from oscillation data, $\theta_{13}=0$ 
has been ruled out experimentally at more than $9\sigma$ confidence level. This has led to a significant 
number of interesting works trying to explain the origin of non-zero $\theta_{13}$. Here we study the 
possibility of explaining the deviations from TBM mixing and hence from $\theta_{13}=0$ by allowing the 
type II seesaw term as a perturbation. It should be noted that the structure of the type I seesaw mass matrix (\ref{matrix1}) does not constrain the Dirac neutrino mass matrix $M_{LR}$ or the right handed neutrino mass matrix $M_{RR}$ to have some specific form. However, choosing one to have some particular form restricts the other so as to get the desired type I seesaw structure (\ref{matrix1}). For example, if we choose the Dirac neutrino mass matrix to have a diagonal structure
\begin{equation}
M_{LR}=\left(\begin{array}{ccc}
a& 0&0\\
0& b & 0 \\
0 & 0 & c
\end{array}\right)
\label{matrixmLR}
\end{equation}
then the $M_{RR}$ is restricted to have the following form 
\begin{equation}
M_{RR}=\left(\begin{array}{ccc}
\frac{a^2(x+y)}{x^2+xy-2y^2}& -\frac{aby}{x^2+xy-2y^2}& -\frac{acy}{x^2+xy-2y^2}\\
-\frac{aby}{x^2+xy-2y^2}& \frac{b^2(x^2-y^2+xz)}{(x^2+xy-2y^2)(x-y+2z)} & \frac{bc(y^2-xy+xz)}{(x^2+xy-2y^2)(x-y+2z)} \\
-\frac{acy}{x^2+xy-2y^2} & \frac{bc(y^2-xy+xz)}{(x^2+xy-2y^2)(x-y+2z)} & \frac{c^2(x^2-y^2+xz)}{(x^2+xy-2y^2)(x-y+2z)}
\end{array}\right)
\label{matrixmRR}
\end{equation}

Before choosing the minimal structure of the type II seesaw term, we note that the parametrization 
of the TBM plus corrected neutrino mass matrix can be done as \cite{nzt13GA}.
\begin{equation}
m_{LL}=\left(\begin{array}{ccc}
x& y-w&y+w\\
y-w& x+z+w & y-z \\
y+w & y-z& x+z-w
\end{array}\right)
\label{matrix2}
\end{equation}
where $w$ denotes the deviation of $m_{LL}$ from that within TBM frameworks and setting it to zero, 
the above matrix boils down to the familiar $\mu-\tau$ symmetric matrix (\ref{matrix1}). Thus, the 
minimal structure of the perturbation term to the leading order $\mu-\tau$ symmetric TBM neutrino 
mass matrix can be taken as
\begin{equation}
m^{II}_{LL}=\left(\begin{array}{ccc}
0& -w& w\\
-w& w & 0 \\
w & 0& -w
\end{array}\right)
\label{matrix3}
\end{equation}
Such a minimal form of the type II seesaw term can be explained by incorporating additional 
flavor symmetries as outlined in \cite{dborah7-13}.

\begin{table}
\centering
\begin{tabular}{|c|c|c|c|c|}
 \hline
   Parameters & IH($m_3 = 0.001$eV) &  IH($m_3 = 0.065$eV) &  NH($m_1 = 0.001$eV)&  NH($m_1=0.07$eV)\\ \hline
x&0.0487942&0.0812709&0.0035726&0.0701779\\  \hline
y&0.0002555&0.0001536&0.0025726&0.0001778\\  \hline
z&-0.023769&-0.0080586&0.0243546&0.007924\\  \hline
$m_3\;(\text{eV})$&0.001&0.065&0.049&0.0858\\  \hline
$m_2\; (\text{eV})$&0.049&0.0815&0.008&0.0705\\  \hline
$m_1\;(\text{eV})$&0.048&0.0811&0.001&0.07\\  \hline
$\sum_i m_i\;(\text{eV})$&0.0988&0.2276&0.0594&0.2263\\  \hline
\end{tabular}
\caption{Parametrization of the neutrino mass matrix for TBM mixing}
\label{table:results1}
\end{table}
We first numerically fit the leading order $\mu-\tau$ symmetric neutrino mass matrix (\ref{matrix1}) 
by taking the central values of the global fit neutrino oscillation data \cite{schwetz12} as presented 
in table \,\ref{table-osc}. We also incorporate the cosmological upper bound on the sum of absolute 
neutrino masses $\sum_i m_{i} < 0.23$ eV \cite{Planck13} reported by the Planck collaboration recently. 
For normal hierarchy, the diagonal mass matrix of the light neutrinos can be written  as $m_{\text{diag}} 
= \text{diag}(m_1, \sqrt{m^2_1+\Delta m_{21}^2}, \sqrt{m_1^2+\Delta m_{31}^2})$ whereas for inverted hierarchy 
 it can be written as $m_{\text{diag}} = \text{diag}(\sqrt{m_3^2+\Delta m_{23}^2-\Delta m_{21}^2}, 
\sqrt{m_3^2+\Delta m_{23}^2}, m_3)$. We choose two possible values of the lightest mass eigenstate $m_1, 
m_3$ for normal and inverted hierarchies respectively. First we choose $m_{\text{lightest}}$ as large 
as possible such that the sum of the absolute neutrino masses fall just below the cosmological upper bound. 
For normal and inverted hierarchies, this turns out to be $0.07$ eV and $0.065$ eV respectively. Then 
we allow moderate hierarchy to exist between the mass eigenvalues and choose the lightest mass eigenvalue 
to be $0.001$ eV to study the possible changes in our analysis and results. The parametrization for all 
these possible cases are shown in table\, \ref{table:results1}. 

\begin{figure}[ht]
 \centering
\includegraphics[width=1.0\textwidth]{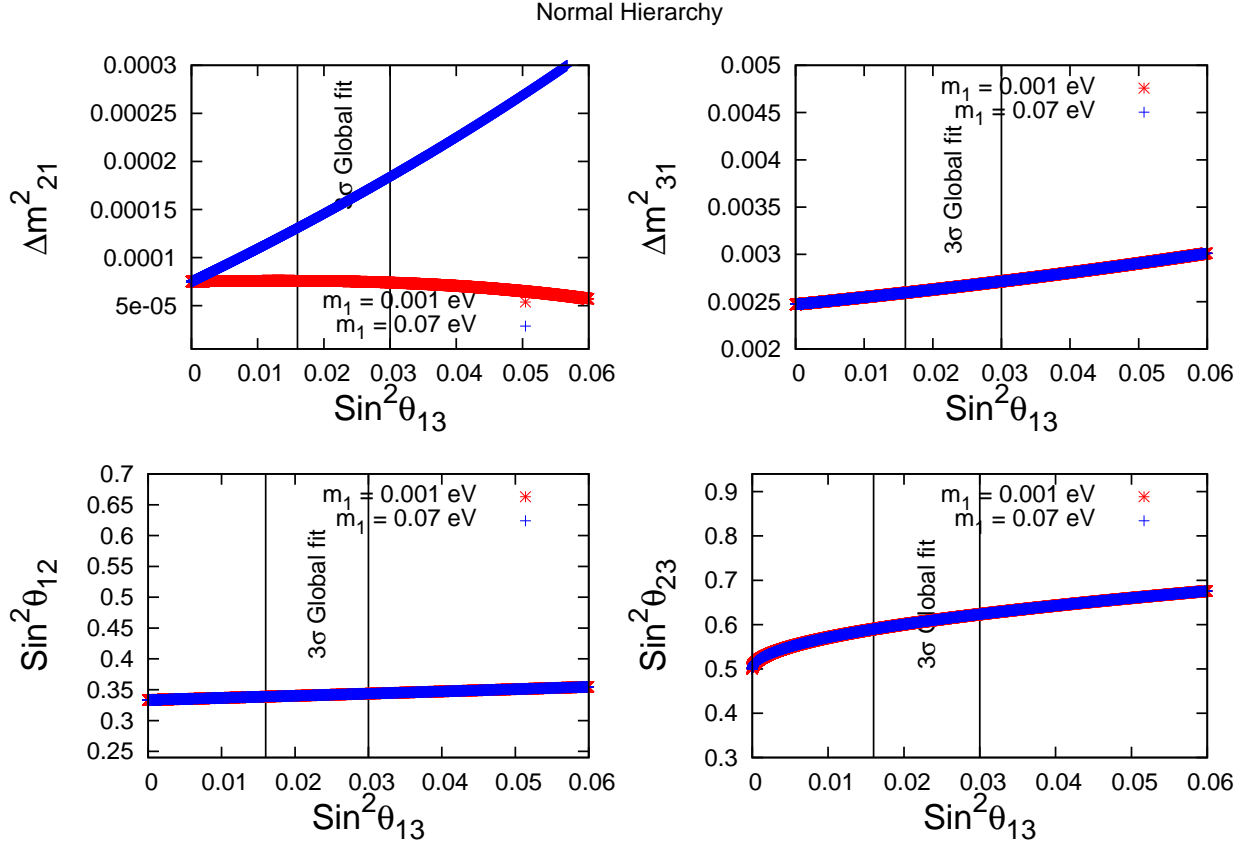}
\caption{Variation of neutrino parameters as a function of $\sin^2{\theta_{13}}$ for Normal Hierarchy}
\label{fig2}
\end{figure}
\begin{figure}[ht]
 \centering
\includegraphics[width=1.0\textwidth]{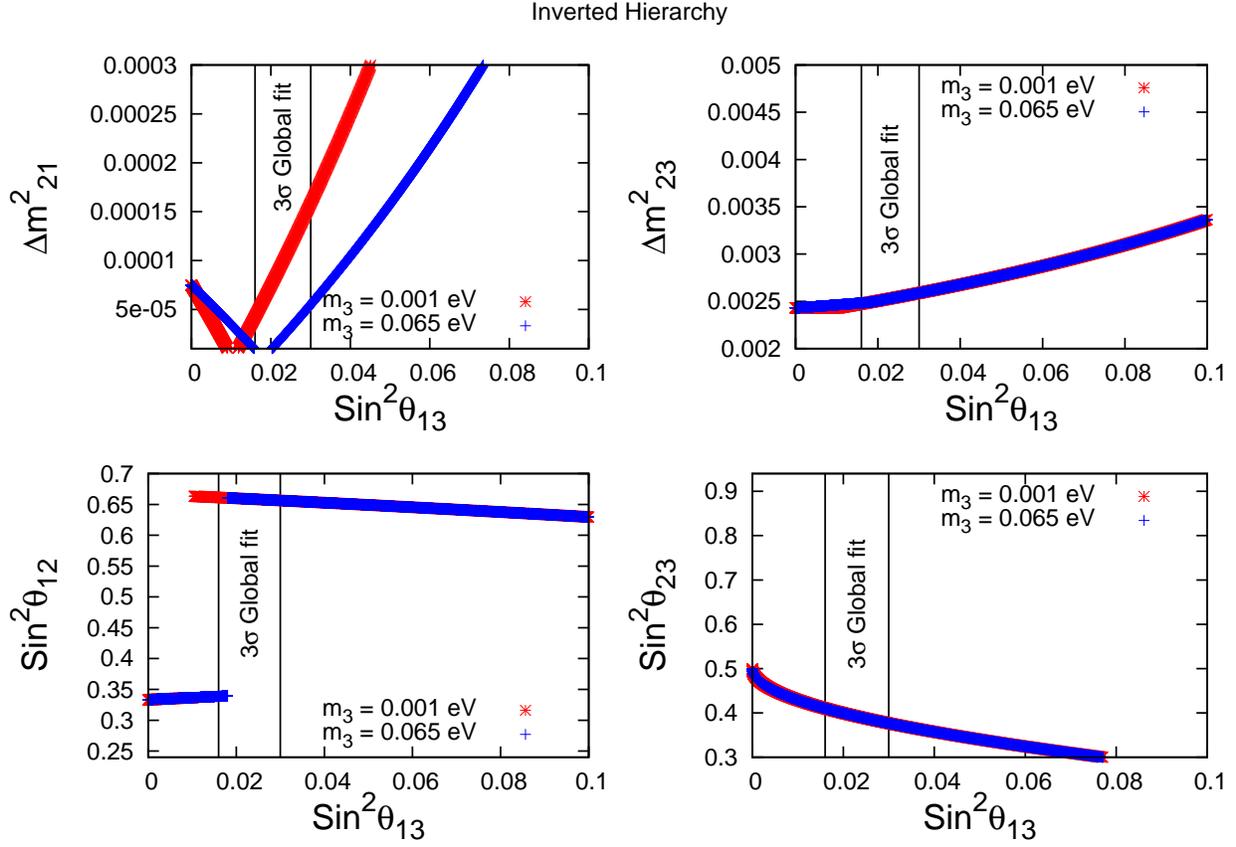}
\caption{Variation of neutrino parameters as a function of $\sin^2{\theta_{13}}$ for Inverted Hierarchy}
\label{fig3}
\end{figure}
\begin{figure}[ht]
 \centering
\includegraphics[width=1.0\textwidth]{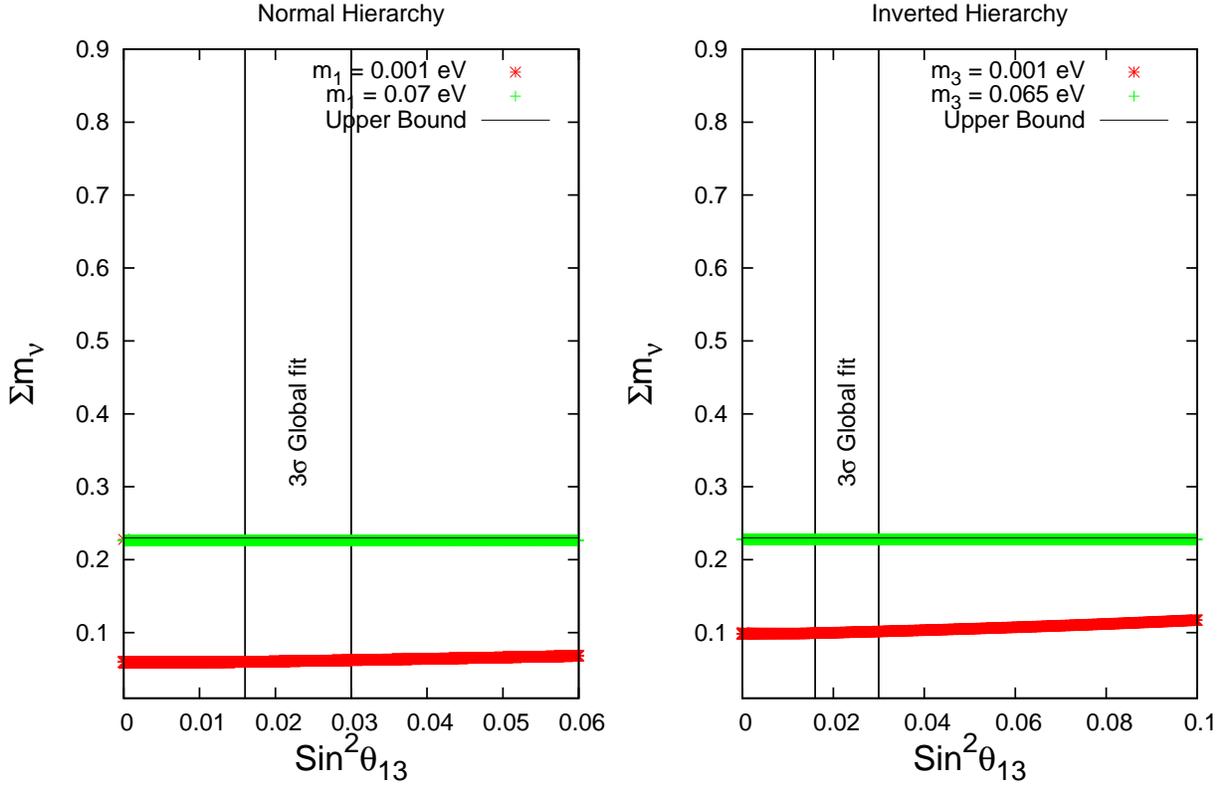}
\caption{Variation of the sum of absolute neutrino masses as a function of $\sin^2{\theta_{13}}$}
\label{fig4}
\end{figure}
\begin{figure}[ht]
 \centering
\includegraphics[width=1.0\textwidth]{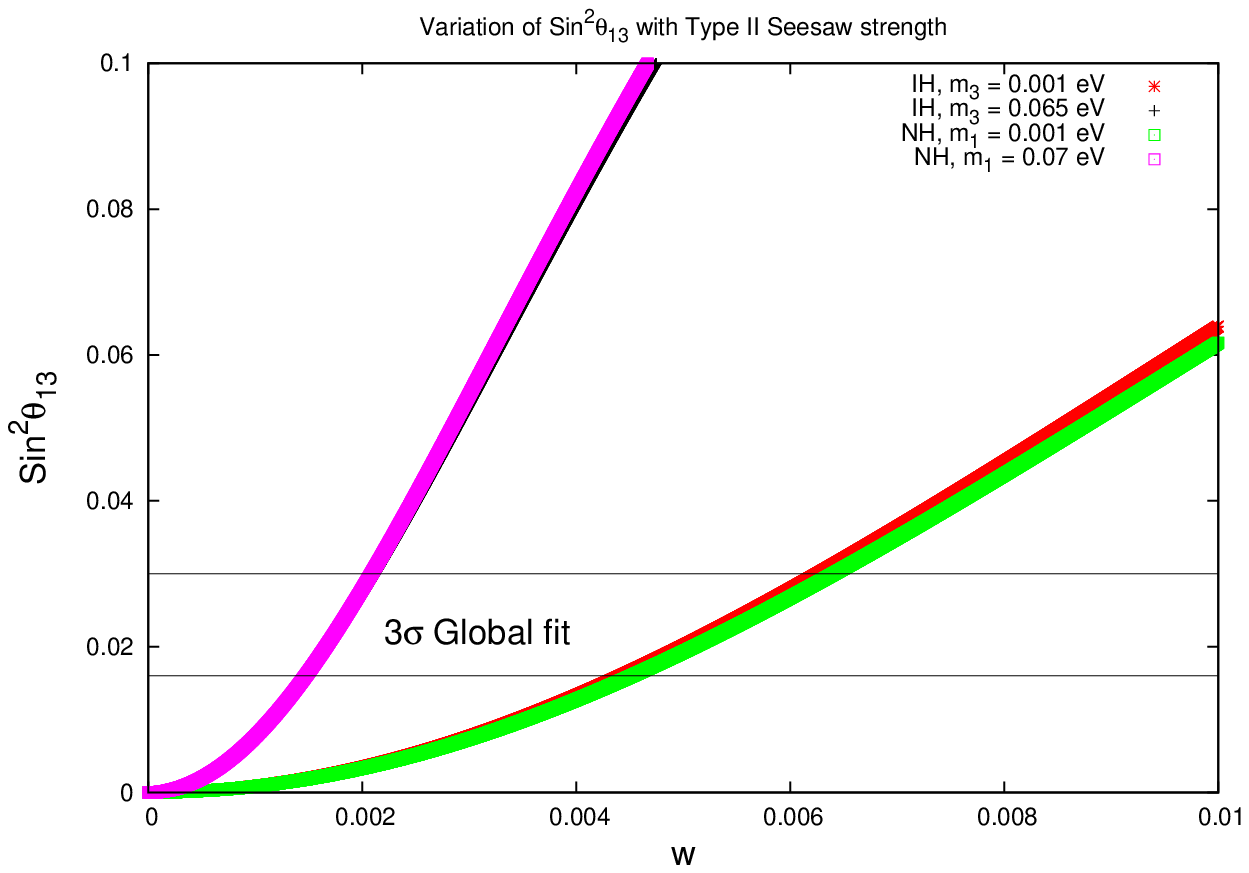}
\caption{Variation of $\sin^2{\theta_{13}}$ as a function of type II seesaw strength}
\label{fig5}
\end{figure}

We then incorporate the type II seesaw contribution which breaks $\mu-\tau$ symmetry and hence gives rise 
to non-zero $\theta_{13}$. We show the variation of other neutrino parameters with respect to $\sin^2{\theta_{13}}$ 
in figure \ref{fig2}, \ref{fig3} for normal and inverted hierarchies respectively. It can be seen that the 
differences in the lightest active neutrino mass show up only in the variation of $\Delta m_{21}^2$. In case 
of normal hierarchy, all the parameters lie in the $3\sigma$ range for $m_{\text{lightest}} = 0.07$ eV whereas 
for inverted hierarchy we see a preference for lighter $m_{\text{lightest}}$ namely, $0.001$ eV. We then show 
the variation of sum of absolute neutrino masses in figure \ref{fig4} and for all the cases considered, the sum 
is found to be within the cosmological limit. We also show the variation of $\sin^2{\theta_{13}}$ as a function 
of type II seesaw strength $w$ in figure \ref{fig5}. It is seen that for higher values of $m_{\text{lightest}}$, 
we require a lower strength of the type II seesaw term to give rise to the desired $\theta_{13}$. For 
$m_{\text{lightest}}=0.065, 0.07$ eV, one can see from figure \ref{fig5} that $w \sim 0.002 \text{eV} 
\Rightarrow f\beta \frac{v^2 v_{R}}{M M_P} = 0.002$ eV. Taking the dimensionless couplings to be of order 
unity and $v = 10^2$ GeV, $v_{R} = 10^4$ GeV, one gets a constraint $M M_P \sim 5 \times 10^{19} \; \text{GeV}^2$.  Similarly, for $m_{\text{lightest}}=0.001$ eV, one can estimate this bound to be around $2 \times 10^{19} \; 
\text{GeV}^2$. Thus, from the constraint of neutrino mass, we get a bound on the $SU(2)_R \times D$ breaking 
scale to be of the order of $10^9-10^{10}$ GeV which is consistent with the gauge coupling unification as will 
be discussed below.

The variation of the neutrino parameters with the perturbation strength can be understood simply 
by calculating the diagonalizing matrix of the neutrino mass matrix considered in the study.
\begin{equation}
m_{LL}=\left(\begin{array}{ccc}
x& y-w&y+w\\
y-w& x+z+w & y-z \\
y+w & y-z& x+z-w
\end{array}\right)
\label{matrix3}
\end{equation}
which has eigenvalues $m_1 = x-y+z -\sqrt{3w^2+z^2}$, $m_2 = x+2y$ and $m_3 = x-y+z+\sqrt{3w^2+z^2}$. 
Assuming $m_1 < m_2 < m_3$ we calculate the neutrino parameters by first identifying the diagonalizing 
matrix. Assuming $w$ to be small such that higher order terms beyond $w^2$ can be neglected, we arrive 
at the following approximate variations of neutrino parameters   
        
\begin{equation}
\sin^2{\theta_{13}} = \frac{w^2}{2z^2}+ \text{h.o.}
\end{equation}
\begin{equation}
\sin^2{\theta_{12}} = \frac{1}{3(1-\frac{w^2}{2z^2})} +\text{h.o.}
\end{equation}
\begin{equation}
\sin^2{\theta_{23}} = \frac{(3y-2z-(1-\frac{3y}{2z})w-\frac{3w^2}{2z})^2}{2(3y-2z)^2} +\text{h.o.}
\end{equation}
\begin{equation}
\Delta m_{21}^2 = (x+2y)^2-(x-y+z-\sqrt{3w^2+z^2})^2
\end{equation}
\begin{equation}
\Delta m_{31}^2 = 4(x-y+z) \sqrt{3w^2+z^2}
\end{equation}
where h.o. refers to higher order terms in $w$. It can be easily seen that for $w=0$, the mixing angles 
correspond to the values predicted by TBM mixing.
\section{Embedding the model in non-SUSY SO(10) GUT}
\label{embedso10}
With the rich phenomenology of the TeV scale asymmetric left-right model discussed in previous sections, we now intend to embed 
the model in a non-supersymmetric SO(10) grand unified theory. We examine whether the model unifies the three gauge couplings successfully with the
proton life time lying close to the experimental lower bound and at the same time allows the possibility of TeV scale $Z^\prime$, RH Majorana neutrinos and RH 
Higgs triplets which can be directly probed at ongoing experiments like LHC. The desired symmetry breaking pattern of SO(10) gauge group with left-right symmetry 
as an intermediate step is given by
{\small 
\begin{eqnarray}
SO(10)& &\hspace*{-0.0cm}\mathop{\longrightarrow}^{M_U}_{} SU(2)_L \times SU(2)_{R} \times U(1)_{B-L} \times SU(3)_C \times D 
                 \quad \left[\mathcal{G}_{2213D}, \, \, (g_{2L} = g_{2R})  \right]\nonumber \\ 
& &\hspace*{-0.0cm} \mathop{\longrightarrow}^{M_P}_{} SU(2)_L \times U(1)_{R} \times U(1)_{B-L} \times SU(3)_C 
              \quad \left[\mathcal{G}_{2213} \,  (g_{2L} \neq g_{2R}) \right]\nonumber \\  
& &\hspace*{-0.0cm} \mathop{\longrightarrow}^{M^0_R}_{} SU(2)_L \times U(1)_{Y} \times SU(3)_C 
              \quad \left[\mathcal{G}_{\rm SM} \equiv \mathcal{G}_{\rm 213}\right] \nonumber \\
& &\hspace*{-0.0cm} \mathop{\longrightarrow}^{M_Z}_{}~U(1)_{\rm em}\times SU(3)_C \quad \quad   \left[\mathcal{G}_{\rm 13}\right]\, .
\label{chain}              
\end{eqnarray}
}
With the above choice of symmetry breaking, the SO(10) gauge group gets broken down to the Standard Model gauge group via the intermediate symmetry 
breaking chain as $\mathcal{G}_{2213D}$, and $\mathcal{G}_{2113}$. The breaking of $SO(10)$ group to LR gauge group is 
achieved by $\{210_H\}$ representation of $SO(10)$ Higgs. The decomposition of $\{210_H\}$ under Pati-Salam gauge group $SU(2)_L\times SU(2)_R \times SU(4)$ is $\Upsilon\{210_H\} \equiv (1, 1, 1) 
\oplus (1, 1, 15) \oplus  (3,1,15) \oplus (1,3,15) \oplus (2, 2, 10) \oplus (2, 2, \bar{10}) \oplus (2, 2, 6)$. The $SO(10)$ 
symmetry can be broken by assigning a VEV to $\langle (1, 1, 15 \rangle$ of $\{210_H\}$ being even under D-parity ensuring 
discrete left-right symmetry (D-parity) intact at this stage. Such a Higgs choice, however, does not affect our mechanism of neutrino mass generation. The second stage of symmetry breaking from $\mathcal{G}_{2213D} \, \, (g_{2L} = g_{2R})$ to $\mathcal{G}_{2113} 
\,  (g_{2L} \neq g_{2R}) $ is done via combination of Higgs representation $\{45\}_H$, and $\{54\}_H$. This is the minimal choice 
of Higgs representation that is necessary to obtain the required symmetry breaking chain consistent with extended survival hypothesis. 
The principle of extended survival hypothesis says that at every stage of symmetry breaking chain we allow only those scalars to be present that 
acquire VEVs at the current or the subsequent levels of spontaneous symmetry breaking. This is equivalent to minimal number 
of fine-tunings to be imposed on the Higgs scalar potential so that all necessary symmetry breaking steps are executed
at the desired scales. Under $\mathcal{G}_{224}$ and $\mathcal{G}_{2213}$, the Higgs representations $\{45\}_H$, and $\{54\}_H$ can be decomposed as 
\begin{eqnarray}
& &\mathcal{S}\{54\}_H \equiv (1, 1, 1) \oplus (3, 3, 1) \oplus (1, 1, 20) \oplus (2, 2, 6) \quad \mbox{under}\quad \mathcal{G}_{224}\, ,\nonumber \\
& &\hspace*{1.3cm} \subset (1, 1, 0, 1) \oplus (3, 3, 0, 1) \oplus (1, 1, 0, 8) \oplus (1, 1, -2/3, 6) \oplus (1, 1, -2/3, \bar{6}) \nonumber \\
& &\hspace*{1.6cm} \oplus (2, 2, 1/3, 3) \oplus (2, 2, -1/3, \bar{3}) \quad \mbox{under}\quad \mathcal{G}_{2213}\, ,
\nonumber \\
& &\mathcal{A}\{45\}_H \equiv (3, 1, 1) \oplus (1, 3, 1) \oplus (2, 2, 6) \oplus (1, 1, 15) 
\quad \mbox{under}\quad \mathcal{G}_{224}\, ,\nonumber
\nonumber \\
& &\hspace*{1.3cm} \subset (1, 1, 0, 1) \oplus \Omega_L(3, 1, 0, 1) \oplus \Omega_R(1, 3, 0, 1) \oplus (2, 2, 1/3, 3) \oplus (2, 2, -1/3, \bar{3}) \nonumber \\
& &\hspace*{1.6cm} \oplus (1, 1, 2/3, 3) \oplus (1, 1, -2/3, \bar{3}) \oplus (1, 1, 0, 8) \quad \mbox{under}\quad \mathcal{G}_{2213}\, .
\label{higgs:decomp}
\end{eqnarray}
The remaining symmetry breaking $SU(2)_L \times U(1)_{R} \times U(1)_{B-L} \times SU(3)_C$ to the SM gauge group $SU(2)_L \times U(1)_{Y} 
\times SU(3)_C $ is implemented by $\{ 126\}_H$ Higgs representation. The decomposition of $\{ 126\}_H$ under Pati-Salam gauge group 
is $\{ 126\}_H \equiv (2, 2, 15) \oplus (3, 1, 10) \oplus (1, 3, \bar{10}) \oplus (1, 1, 6)$. Assigning a VEV to $\langle \Delta_R(1, 1, -2, 1) \rangle 
\subset \Delta_R(1, 3, -2, 1) \subset (1, 3, \bar{10})$, we break $U(1)_{R} \times U(1)_{B-L}$ to $U(1)_{Y}$. The last stage of symmetry 
breaking of the SM gauge group $SU(2)_L \times U(1)_{Y} \times SU(3)_C$  to $U(1)_{\rm em} \times SU(3)_C$ is achieved by $\{10_H\}$ where the 
Higgs field $\Phi(2, 1/2, 1) \subset (2, 2, 0, 1) \subset \{10_H\}$ acquires a VEV breaking $SU(2)_L \times U(1)_{Y}$  to $U(1)_{\rm em}$.
In the following sections, we present the gauge coupling evolution with the evaluation of one-loop beta coefficients and estimate the 
proton life time $\tau_p$ using the value of gauge coupling at GUT scale.
\section{Gauge coupling evolution with one-loop analysis}
\label{unifso10}
In this section we study the one loop renormalization group evolution (RGE) equations for gauge couplings relevant for our model. The one loop RGE equations for the gauge couplings can be written as
\begin{eqnarray}
\frac{d\, \alpha^{-1}_{i}}{d\, t}=-\frac{\pmb{a_i}}{2 \pi}
\label{rge-coupl}
\end{eqnarray}
where $t=\ln(\mu)$, $\alpha_{i}=g^2_{i}/(4 \pi)$ are the fine structure constants and $\pmb{a_i}$ are the one-loop beta coefficients 
derived for the corresponding $i^{\rm th}$ gauge group for which coupling evolution has to be determined. The analytic formula 
for $\pmb{a_i}$ is
\begin{eqnarray}
\pmb{a_i}= -\frac{11}{3} \mathcal{C}_{2}(\mathcal{G}_i) + \frac{4}{3} \kappa\, N_G + \frac{1}{3} \eta T(R_{S_i})\, d(S_i) \, ,
\label{beta:coeff}
\end{eqnarray}
with no summation over $i$. We denote $\mathcal{C}_{2}$ and $\mathcal{T}_{2}$ as quadratic Casimir of a given representation, $d_{S_i}$ 
as the multiplicity factor for a particular gauge group $\mathcal{G}_i$ due to other $SU(N)_j$ group present in the model, $N_G$ as the number 
of fermion generation (which is 3 in our model). We take $\kappa=1, 
\frac{1}{2}$ for Dirac and Weyl fermions, $\eta=1, \frac{1}{2}$ for complex and real scalar fields, respectively. 
\subsection{Matching condition and estimations for $M_U$, $M_P$ and $\alpha_U$}
One can write the RGE equations for the standard model gauge couplings in terms of present non-SUSY $SO(10)$ GUT coupling. Since the model 
has two intermediate symmetry breaking steps above standard model scale, it is important to know the appropriate matching condition 
at these two symmetry breaking steps. Denoting $\alpha^{-1}_i = \frac{4 \pi}{g^2_i}$, the appropriate matching conditions for 
gauge couplings valid at the gauge group $\mathcal{G}_{2113} =SU(2)_L \times U(1)_R \times U(1)_{B-L} \times SU(3)_C$ are
{\small \begin{eqnarray}
\hspace*{-0.80cm}\mbox{At}\, \mu=M^0_R:&\quad&
\bigg[\alpha^{-1}_{Y}(M^0_R)\bigg]_{\mathcal{G}_{SM}} = 
       \bigg[ \frac{3}{5} \alpha^{-1}_{1R}(M^0_R) + \frac{2}{5} \alpha^{-1}_{B-L}(M^0_R)\bigg]_{\mathcal{G}_{2113}}\, ,
\nonumber \\
& & \bigg[\alpha^{-1}_{2L}(M^0_R)\bigg]_{\mathcal{G}_{SM}} = \bigg[\alpha^{-1}_{2L}(M^0_R)\bigg]_{\mathcal{G}_{2113}}\, ,
\bigg[\alpha^{-1}_{3C}(M^0_R)\bigg]_{\mathcal{G}_{SM}} = \bigg[\alpha^{-1}_{3C}(M^0_R)\bigg]_{\mathcal{G}_{2113}}\, .
\label{match:mr0}
\end{eqnarray}
}
Similarly, the appropriate gauge coupling matching conditions at the scale $M_P$ valid for the gauge group $\mathcal{G}_{2213D} 
=SU(2)_L \times SU(2)_R \times U(1)_{B-L} \times SU(3)_C \times D$ are 
{\small \begin{eqnarray}
\hspace*{-0.2cm}\mbox{At}\, \mu=M_P&:&
\bigg[\alpha^{-1}_{2L}(M_P)\bigg]_{\mathcal{G}_{2113}} = \bigg[\alpha^{-1}_{2L}(M_P) \bigg]_{\mathcal{G}_{2213D}} \nonumber \\
& &\bigg[\alpha^{-1}_{1R}(M_P)\bigg]_{\mathcal{G}_{2113}} = \bigg[\alpha^{-1}_{2R}(M_P)\bigg]_{\mathcal{G}_{2213D}}\, ,
\bigg[\alpha^{-1}_{B-L}(M_P)\bigg]_{\mathcal{G}_{2113}} = \bigg[\alpha^{-1}_{B-L}(M_P)\bigg]_{\mathcal{G}_{2213D}}\, ,
\nonumber \\
& &\bigg[\alpha^{-1}_{3C}(M_P)\bigg]_{\mathcal{G}_{2113}} = \bigg[\alpha^{-1}_{3C}(M_P)\bigg]_{\mathcal{G}_{2213D}}\, ,
\bigg[\alpha^{-1}_{2L}(M_P)\bigg]_{\mathcal{G}_{2113}} = \bigg[\alpha^{-1}_{2R}(M_P)\bigg]_{\mathcal{G}_{2213D}}\, .
\label{match:mp}
\end{eqnarray}
}
Also, one can write down the gauge coupling matching conditions at the unification scale $M_U$ as 
{\small \begin{eqnarray}
\hspace*{-0.2cm}\mbox{At}\, \mu=M_U:&\quad&
\bigg[\alpha^{-1}_{2L}(M_U)\bigg]_{\mathcal{G}_{2213D}} \equiv \bigg[\alpha^{-1}_{2R}(M_U)\bigg]_{\mathcal{G}_{2213}} 
       = \bigg[\alpha^{-1}_{10}(M_U) \bigg]_{SO_{10}}\, , \nonumber \\
& &\bigg[\alpha^{-1}_{B-L}(M_U)\bigg]_{\mathcal{G}_{2213D}} 
       = \bigg[\alpha^{-1}_{10}(M_U) \bigg]_{SO_{10}}\, , \quad 
\bigg[\alpha^{-1}_{3C}(M_P)\bigg]_{\mathcal{G}_{2213D}} 
      = \bigg[\alpha^{-1}_{10}(M_U) \bigg]_{SO_{10}}\, .
\label{match:mu}
\end{eqnarray}
}
With the above gauge coupling matching conditions, one can express the RGE equations for $\alpha^{-1}_i,\, \mbox{i=2L, Y, 3C for SM}$ 
valid at one-loop level
{\small \begin{eqnarray}
& &\alpha^{-1}_{2L} (M_Z)=\alpha^{-1}_{10} (M_U) + \frac{\pmb a_{2L}}{2 \pi} {\large \ln}\left(\frac{M^0_R}{M_Z}\right) 
          + \frac{\pmb a^\prime_{2L}}{2 \pi} {\large \ln} \left(\frac{M_P}{M^0_R}\right) 
          + \frac{\pmb a^{\prime \prime}_{2L}}{2 \pi} {\large \ln} \left(\frac{M_U}{M_P}\right)\, , 
   \label{rge:2L}       \\
& &\alpha^{-1}_{Y} (M_Z)=\alpha^{-1}_{10} (M_U) + \frac{\pmb a_Y}{2 \pi} {\large \ln}\left(\frac{M^0_R}{M_Z}\right) 
          + \frac{\frac{3}{5} \pmb a^\prime_{1R} + \frac{2}{5} \pmb a^\prime_{B-L}}{2 \pi} 
            {\large \ln} \left(\frac{M_P}{M^0_R}\right) \nonumber \\
         & &\hspace*{7cm} +\frac{\frac{3}{5} \pmb a^{\prime \prime}_{2R} + \frac{2}{5} \pmb a^{\prime \prime}_{B-L}}{2 \pi}  
            {\large \ln} \left(\frac{M_U}{M_P}\right)\, , 
  \label{rge:Y}           \\          
& &\alpha^{-1}_{3C} (M_Z)=\alpha^{-1}_{10} (M_U) + \frac{\pmb a_{3C}}{2 \pi} {\large \ln}\left(\frac{M^0_R}{M_Z}\right) 
          + \frac{\pmb a^\prime_{3C}}{2 \pi} {\large \ln} \left(\frac{M_P}{M^0_R}\right) 
          + \frac{\pmb a^{\prime \prime}_{3C}}{2 \pi} {\large \ln} \left(\frac{M_U}{M_P}\right)\, ,
   \label{rge:3C}        
\end{eqnarray}
}
where the one-loop beta coefficients for our model determined by the particle spectrum in the mass ranges $M_Z - M^0_R$, $M^0_R - M_P$ 
and $M_P - M_U$ are $\{\pmb a_{2L}, \pmb a_{Y}, \pmb a_{3C}\}$, $\{\pmb a^\prime_{2L}, \pmb a^\prime_{1R}, 
\pmb a^\prime_{B-L}, \pmb a^\prime_{3C}\}$, and $\{\pmb a^{\prime \prime}_{2L}, \pmb a^{\prime \prime}_{2R}, \pmb a^{\prime \prime}_{B-L}, 
\pmb a^{\prime \prime}_{3C} \}$, for gauge groups $\mathcal{G}_{213}$, $\mathcal{G}_{2113}$ and $\mathcal{G}_{2213D}$, 
respectively. Fixing $M^0_R$ around few TeV, and using particle data group values \cite{PDG} $\sin^2\theta_W= 0.23166 \pm 0.00005$, $\alpha_S=0.1184 \pm 0.003$, 
and $\alpha_{em}=1/127.94$, a simple one-loop analytical survey of the gauge coupling running equations yields two important relations for $M_P$ 
and $M_U$ as \cite{utpal-desai}
\begin{eqnarray}
& &\mbox{ln}\left( \frac{M_U}{M_Z}\right)=\frac{\mathcal{D}_1 \mathcal{A}_P-\mathcal{D}_0 \mathcal{B}_P}
                {\mathcal{B}_U \mathcal{A}_P-\mathcal{A}_U \mathcal{B}_P}\, , 
                \label{sol:mu} \\
& &\mbox{ln}\left( \frac{M_P}{M_Z}\right)=\frac{\mathcal{D}_0 \mathcal{B}_U-\mathcal{D}_1 \mathcal{A}_U}
                {\mathcal{B}_U \mathcal{A}_P-\mathcal{A}_U \mathcal{B}_P} \, ,  
                \label{sol:mp}
\end{eqnarray}
with \begin{eqnarray*}
& &\mathcal{A}_0=\left( 8 \pmb{a}_{3C}- 3 \pmb{a}_{2L}- 5\pmb{a}_Y \right) 
                 -\left(8 \pmb{a}^\prime_{3C}- 3 \pmb{a}^\prime_{2L}-3 \pmb{a}^\prime_{1R}-2 \pmb{a}^\prime_{B-L}\right),
                 \label{eq1} \\
& &\mathcal{A}_P=\left(8 \pmb{a}^\prime_{3C}- 3 \pmb{a}^\prime_{2L}-3 \pmb{a}^\prime_{1R}-2 \pmb{a}^\prime_{B-L}\right)
                 -\left(8 \pmb{a}^{\prime \prime}_{3C} - 6 \pmb{a}^{\prime \prime}_{2L}- 2 \pmb{a}^{\prime \prime}_{B-L} \right)  \, ,  
                 \label{eq2} \\
& &\mathcal{A}_U= \left(8 \pmb{a}^{\prime \prime}_{3C} - 
                6 \pmb{a}^{\prime \prime}_{2L}- 2 \pmb{a}^{\prime \prime}_{B-L} \right) \, , 
                \label{eq3} \\
& &\mathcal{B}_0= \left(5\pmb{a}_{2L}- 5\pmb{a}_Y \right) 
        - \left(5 \pmb{a}^\prime_{2L}-3 \pmb{a}^\prime_{1R} - 2 \pmb{a}^\prime_{B-L} \right)  \, , 
        \label{eq4} \\
& &\mathcal{B}_P= \left(5 \pmb{a}^\prime_{2L}-3 \pmb{a}^\prime_{1R} - 2 \pmb{a}^\prime_{B-L} \right)
        - \left( 2 \pmb{a}^{\prime \prime}_{2L} - 2 \pmb{a}^{\prime \prime}_{B-L} \right)  \, , 
        \label{eq5} \\
& &\mathcal{B}_U = \left( 2 \pmb{a}^{\prime \prime}_{2L} - 2 \pmb{a}^{\prime \prime}_{B-L} \right)\, , 
          \label{eq6} \\
& &\mathcal{D}_0=16 \pi \left(\alpha^{-1}_s -\frac{3}{8} \alpha^{-1}_{\rm em} \right) 
            - \mathcal{A}_0 \mbox{ln}\left( \frac{M^0_R}{M_Z}\right)\, , 
          \label{eq7} \\
& &\mathcal{D}_1= \frac{16 \pi}{\alpha_{em}} \left(\sin^2 \theta_W -\frac{3}{8} \right) 
            - \mathcal{B}_0 \mbox{ln}\left( \frac{M^0_R}{M_Z}\right)\, .
            \label{eq8}
\end{eqnarray*}
In the following subsection, the value of the D-parity breaking scale $M_P$ and the unification scale $M_U$ 
are estimated using the above model parameters by fixing the $U(1)_R \times U(1)_{B-L}$ breaking 
scale $M^0_R$ around 1 TeV to 6 TeV. The estimation of $M_P$, $M_U$ and $\alpha_U$ following from 
eqn.(\ref{rge:2L}) to eqn.(\ref{sol:mp}) is carried out for different scenarios defined by the 
spectrum of Higgs fields utilized for the purpose of symmetry breaking.

\begin{center}
\hspace*{1cm}
\begin{table}[h!]
\begin{tabular}{|c|c|c|c|}
\hline
Group $G_{I}$ & \mbox{Range of Masses (\mbox{GeV})}       & Higgs content    & $\pmb {a}$         \\
\hline \hline
${\small G_{2_L 1_Y 3_C}}$  & ${\small \bf M_Z - M^0_R}$   & 
$\begin{array}{l}
\Phi(2, \frac{1}{2}, 1)_{10}\end{array}$
                               & 
$\pmb {a}_i$=$\bmt 
-19/6 \\
41/10 \\
-7
\emt$  \\
\hline 
${\small G_{2_L 1_R 1_{B-L} 3_C}}$   & ${\small \bf M^0_R - M_P}$ &  
${\small \begin{array}{l}
\Phi_1(2,\frac{1}{2},0,1)_{10} \oplus \Phi_2(2,-\frac{1}{2},0,1)_{10^\prime}\\
\oplus \Delta_R(1,1,-2,1)_{126} \end{array}}$   
                                                                                          &
$\pmb{a}^{\prime}_i$= $\bmt 
-3 \\ 
14/3\\
9/2\\
-7
\emt$                                                                                      
\\
\hline 
${\small G_{2_L 2_R 1_{B-L} 3_C D}}$  & ${\small \bf M_P- M_U}$  & 
${\small \begin{array}{l}
\Phi_1(2,2,0,1)_{10}\oplus \Phi_2(2,2,0,1)_{10^\prime} \\
\oplus \Delta_R(1,3,-2,1)_{126} \oplus \Delta_L(3,1,-2,1)_{126} \\
\oplus \Sigma_R(1,3,0,1)_{210} \oplus \Sigma_L(3,1,0,1)_{210} 
\end{array}}$
                                         &
$\pmb{a}^{\prime \prime}_i$ = $\bmt -4/3 \\
      -4/3 \\
       7\\
      -7 \emt$ \\
\hline
\end{tabular}
\caption{One-loop beta coefficients for different gauge coupling evolutions. The allowed range of mass scales are 
          $M_Z=91.187$ GeV, $M_R^0=3-6$ TeV, $M_P=1.6 \times 10^{11}$ GeV, and $M_U=1.2 \times 10^{15}$.}
\label{tab:beta_coeff-a}
\end{table}
\end{center}

\noindent
{\bf Breaking of $U(1)_R \times U(1)_{B-L} \to U(1)_Y$ via only Higgs triplet}

\noindent
We note that the contributions to one-loop beta coefficients coming from the fermion and gauge sector are well known and simple 
for a given gauge group while the Higgs contributions to beta-coefficients are complicated due to various Higgs fields present in our 
model. The economical choice of Higgs spectrum for different mass ranges is presented in table \, \ref{tab:beta_coeff-a}.
We find the D-parity breaking scale $M_P$ and the unification scale $M_U$ for the set of 
one-loop beta coefficients given in table \, \ref{tab:beta_coeff-a} to be $M_P=1.6 \times 10^{11}$ GeV and $M_U= 
1.2 \times 10^{15}$ GeV. The above calculated value of $M_U$ results in predicting proton life time $1.2 \times 10^{33}$ 
yrs while the current experimental bound on proton life time is $> 8.2 \times 10^{33}$ yrs. Therefore, it is 
important to discuss the GUT threshold corrections to this unification mass scale in order to know how far 
we are from the experimental lower bound on proton life time. However, we do not perform such an exercise of calculating GUT threshold corrections in this work.

Alternatively, one can try to check the gauge coupling unification with higher unification scale by incorporating 
the presence of additional Higgs fields at different stages of symmetry breaking allowed in the model. With this motivation, 
we include extra Higgs fields $\zeta (1,0,8)$ and $\xi(2,1/2,8)$ (with SM quantum numbers shown within brackets) to 
the minimal particle content of table \,\ref{tab:beta_coeff-a} and numerical values of $M_P$, $M_U$, and $\alpha^{-1}_U$ 
are estimated in the following paragraph. 

\begin{table}[h!]
\centering
\begin{tabular}{|c|c|}
\hline 
{\bf C1:} & {\bf C2:} \\
\hline \hline 
$\left(-19/6, 41/10, -7\right)$                    & $\left(-19/6, 41/10, -7\right)$ 
\\ 
$\left(-3, 14/3, 9/2, -6\right)$            & $\left(-3, 14/3, 9/2, -7\right)$  
\\ 
$\left(-4/3, -4/3, 7, -6\right)$ & $\left(4/3, 4/3, 7, -3\right)$  
\\
\hline
\end{tabular}
\caption{Calculated values of one-loop beta coefficients presented by adding an extra Higgs fields 
to the minimal Higgs content given in table \,\ref{tab:beta_coeff-a}. The one-loop beta coefficients 
are presented as $\pmb{a}_i$, $\pmb{a}^\prime_i$, and $\pmb{a}^{\prime \prime}_i$ in 1st, 2nd and 3rd 
row of each column, respectively. The allowed range of mass scales are 
          $M_Z=91.187$ GeV, $M_R^0=3-6$ TeV, $M_P=10^{9}-10^{11}$ GeV, and $M_U=10^{14.5}-10^{16.5}$}
\label{tab:input-betacoeff-I}
\end{table} 

\begin{figure}[h!]
\centering
\includegraphics[width=0.75\textwidth]{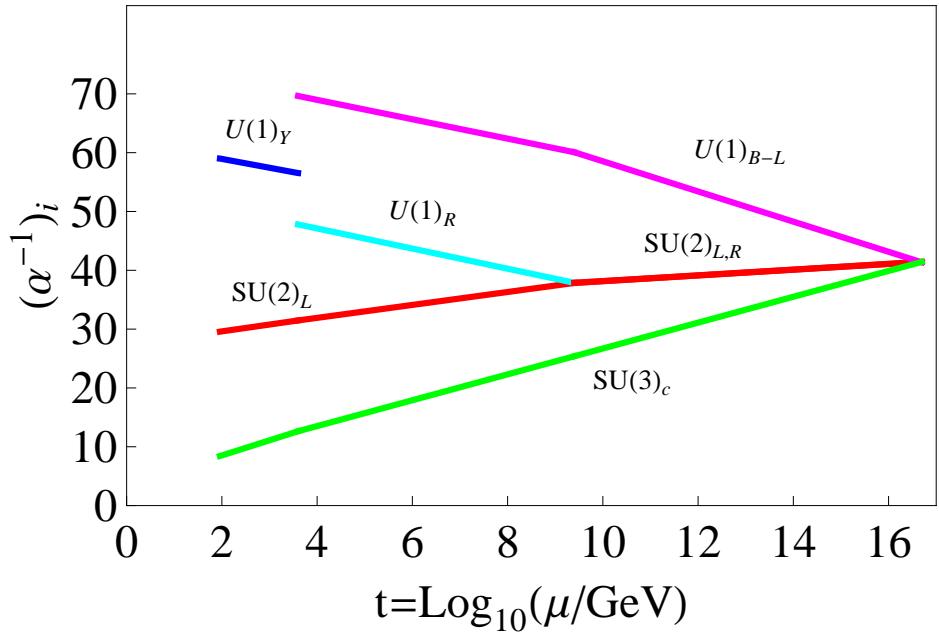}
\caption{One-loop gauge coupling evolution for left-right model with beta functions given in column C1 of table\, \ref{tab:input-betacoeff-I}}
\label{fig:coupl-unifna}
\end{figure}

For evaluation of $\pmb{a_i}$, $\pmb{a^\prime_i}$, and $\pmb{a}^{\prime \prime}_i$ presented under column ${\bf C1}$ of table \,\ref{tab:input-betacoeff-I}, 
the Higgs field $\zeta (1,1,0,8)$ (with $\mathcal{G}_{2213}$ quantum numbers shown within brackets) is added at or above 
the symmetry breaking scale $M^0_R$. The gauge coupling unification for such a case is shown in figure \ref{fig:coupl-unifna}. Similarly, for the evaluation of $\pmb{a_i}$, $\pmb{a}^\prime_i$, and $\pmb{a}^{\prime \prime}_i$ 
presented under column ${\bf C2}$ of table \,\ref{tab:input-betacoeff-I}, the Higgs field $\xi (2,2,0,8)$ is introduced at or above the scale $M_P$. 

\begin{center}
\hspace*{2cm}
\begin{table}[htb!]
\begin{tabular}{|c|c|c|c|c|}
\hline
Higgs content for      & $M^0_R (\mbox{in GeV})$   & $M_P (\mbox{in GeV})$   & $M_U (\mbox{in GeV})$   & $\alpha^{-1}_U$    \\
\hline \hline
{\bf For Table-\ref{tab:beta_coeff-a}}       & \mbox{(3-6) TeV}      & $1.65 \times 10^{11}$   & $1.2 \times 10^{15}$    & $40.9827$      \\
\hline
{\bf For Table-\ref{tab:input-betacoeff-I}: C1}  & \mbox{(3-6) TeV}  & $2.6 \times 10^{9}$    & $4.9 \times 10^{16}$    & $40.7687$      \\
\hline
{\bf For Table-\ref{tab:input-betacoeff-I}: C2}  & \mbox{(3-6) TeV}  & $1.38 \times 10^{11}$    & $1.1 \times 10^{16}$    & $37.946$      \\ 
\hline \hline 
\end{tabular}
\caption{Allowed solutions for different mass scales, and inverse fine structure constant ($\alpha^{-1}_U$) 
        at unification scale consistent with gauge coupling unification.}
\label{tab:pred:masses-model-I}
\end{table}
\end{center}

\noindent
{\bf Breaking of $U(1)_R \times U(1)_{B-L} \to U(1)_Y$ via Higgs triplet ($\Delta$) plus Higgs doublet ($\chi$)} 

\noindent
It should be noted that, shifting the parity breaking scale $M_P$ towards the GUT scale provides us with more 
possibilities to achieve unification with more minimal set of additional fields than discussed above. However, 
to keep a sizable contribution of type II seesaw so that it can give rise to the observed $\theta_{13}$, we intend 
to keep $M_P$ as low as $10^9-10^{10}$ GeV. Here lies the need to include additional field content discussed 
in previous subsection. Apart from the scenario where $U(1)_R \times U(1)_{B-L}$ gauge symmetry is broken 
by Higgs triplets, there can be one more possibility to achieve the same using both triplets and doublets. 
For the sake of completeness we discuss this case as well and check the gauge coupling unification.

In such a scenario, we allow the breakdown of the intermediate symmetry $SU(2)_L \times U(1)_R \times 
U(1)_{B-L} \times SU(3)_C \to SU(2)_L \times U(1)_Y \times SU(3)_C$ driven by both triplet $\Delta$ 
and doublet $\chi$ coming from $126_H$ and $16_H$ representation of $SO(10)$ respectively. The relevant 
Higgs spectrum and the corresponding one-loop beta coefficients for $SO(10) \to \mathcal{G}_{2213D} 
\to \mathcal{G}_{2113} \to \mbox{SM}$ are listed in table \,\ref{tab:beta_coeff-b}. The predicted values 
of the mass scales for this ranges of input parameters are $M_P (\mbox{in GeV})=1.51 \times 10^{11}$, 
$M_U (\mbox{in GeV})=1.02 \times 10^{15}$, and $\alpha^{-1}_U=42.02$.
 
\begin{center}
\hspace*{2cm}
\begin{table}[htb!]
\begin{tabular}{|c|c|c|}
\hline
Group $G_{I}$        & Higgs content    & $\pmb { a_i}$         \\
\hline \hline
${\small G_{2_L 1_Y 3_C}}$                & 
$\begin{array}{l}
\Phi(2, \frac{1}{2}, 1)_{10}\end{array}$
                               & 
$\pmb{a}_i$ = $\bmt 
-19/6 \\
41/10 \\
-7
\emt$  \\
\hline 
${\small G_{2_L 1_R 1_{B-L} 3_C}}$                                                                  &  
${\small \begin{array}{l}
\Phi_1(2,\frac{1}{2},0,1)_{10}\oplus \Phi_2(2,-\frac{1}{2},0,1)_{10^\prime}\\
\Delta_R(1,1,-2,1)_{126} \oplus \chi_R(1,\frac{1}{2},-1,1)_{16} \end{array}}$   
                                                                                          &
$\pmb{a}^{\prime}_i$ = $\bmt 
-3 \\ 
19/4\\
37/8\\
-7
\emt$                                                                                      
\\
\hline 
${\small G_{2_L 2_R 1_{B-L} 3_C D}}$                          & 
${\small \begin{array}{l}
\Phi_1(2,2,0,1)_{10}\oplus \Phi_2(2,2,0,1)_{10^\prime}+ \Delta_R(1,3,-2,1)_{126}\oplus\\
\Delta_L(3,1,-2,1)_{126} \oplus \chi_R(1,2,-1,1)_{16}\oplus \chi_L(2,1,-1,1)_{16}\\
\oplus \Sigma_R(1,3,0,1)_{210} \oplus \Sigma_L(3,1,0,1)_{210}
\end{array}}$
                                         &
$\pmb{a}^{\prime \prime}_i$ = $\bmt -7/6 \\
      -7/6 \\
       15/2\\
      -7 \emt$ \\
\hline
\end{tabular}
\caption{The estimated one-loop beta coefficients for different gauge coupling evolutions 
         with Higgs the fields relevant for different stages of symmetry breaking. The allowed 
         range of mass scales are $M_Z=91.187$ GeV, $M_R^0=3-6$ TeV, $M_P=1.5 \times 10^{11}$ GeV, 
         and $M_U=1.02 \times 10^{16.5}$}
\label{tab:beta_coeff-b}
\end{table}
\end{center}

\begin{figure}[h!]
\centering
\includegraphics[width=0.75\textwidth]{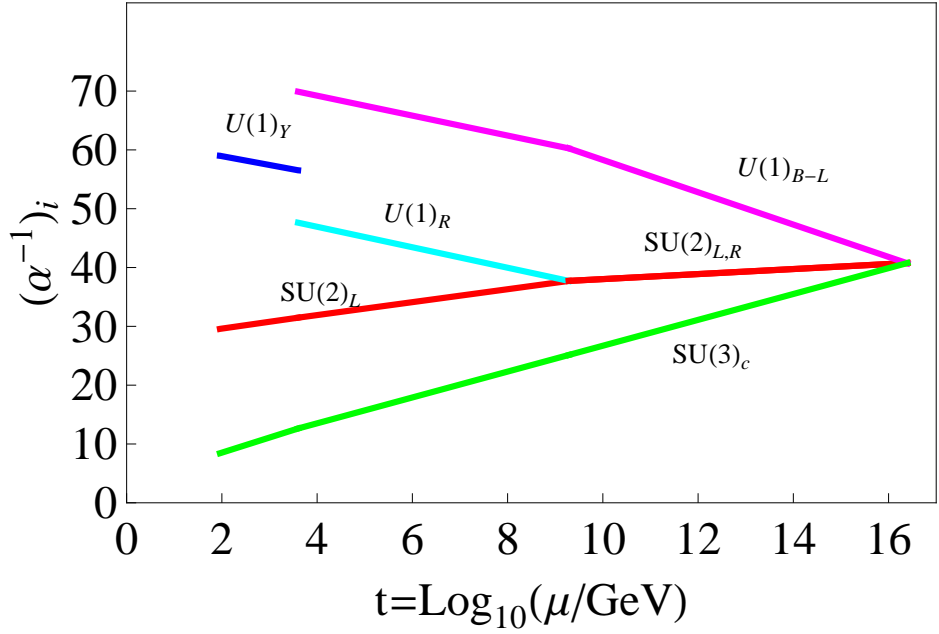}
\caption{One-loop gauge coupling evolution for left-right model with particle content given in table \,\ref{tab:beta_coeff-b} plus an additional color octet Higgs from the scale $M^0_R$ onwards}
\label{fig:coupl-unifnb}
\end{figure}

With addition of extra color octet scalar $\zeta (1,1,0,8)$ from $M^0_R$ onwards relevant for symmetry breaking, 
the derived values of one-loop beta-coefficients are $\pmb{a_i}=\left(-19/6, 41/10, -7\right)$, $\pmb{a}^\prime_i=
\left(-3, 19/4, 37/8, -6\right)$, and $\pmb{a}^{\prime \prime}_i=\left(-7/6, -7/6, 15/2, -6\right)$. As a result, 
the numerically estimated values of mass scales are $M_P (\mbox{in GeV})=1.9 \times 10^{9}$, $M_U (\mbox{in GeV})=2.49 
\times 10^{16}$, and $\alpha^{-1}_U=41.4236$. The coupling evolution for this case is shown in figure \ref{fig:coupl-unifnb}.

It is worth mentioning here that the effect of two-loop RG analysis on gauge coupling 
unification might change the mass scale like $M^0_R \simeq v_R$, $M_P$ and $M_U$. It is found 
that the two-loop RG evolution in this particular non-SUSY $SO(10)$ set up having two intermediate 
symmetry breaking steps changes marginally the values of $M_P$ and $M_U$ as compared to the numerical 
values derived by one-loop RG analysis. We can take the example of two loop analysis having Higgs 
spectrum as presented in Table.VII along with color octet Higgs scalar (1,1,0,8) where the predicted 
mass scales are 
$$M^0_R=v_R\simeq \mbox{(3-10)\, TeV}, M_P\simeq 10^{8.9}\, \mbox{GeV}, M_U \simeq 10^{16.57}\, \mbox{GeV}\, $$
but the findings for one-loop analysis are $$M_P (\mbox{in GeV})=1.9 \times 10^{9}, M_U (\mbox{in GeV})=2.49 
\times 10^{16}\, .$$ Hence, there will be little modification to the type II seesaw contribution which is 
$m^{II}_{\nu}=f v_L=f\,\frac{\beta v^2 v_R}{M M_P}$ if one incudes two-loop RG effect. One can fix the 
neutrino mass arising from type-II seesaw by suitably adjusting the other free parameters like Higgs 
coupling $\beta$ and $M$ even if we include the effect of two-loop RG corrections on $v_R$ and $M_P$.

It should be noted that the LRSM where the breaking of $U(1)_R \times U(1)_{B-L} \to U(1)_Y$ occurs through 
Higgs triplet ($\Delta$) and $SU(2)_R \times D$ gets broken by Higgs triplet $\Sigma$ can also be constrained 
from the cosmologial constraints on the successful disappearance of domain walls. Domain walls generically 
arise in such models (due to the spontaneous breaking of discrete symmetry called D-parity) which, if stable, 
can overclose the Universe conflicting with standard cosmology. As discussed in \cite{borahdwall}, for $M^0_R 
= 10$ TeV, domain wall disappearance requires $M_P < 10^9$ GeV, which are very close to the symmetry breaking 
scales in our present model. Similar constraint on the second model (the one with both Higgs triplet and 
doublet) have not been studied yet and left for future investigations.

\section{Estimation of proton life time $\tau_p$}
\label{pdecay}
With the knowledge of unification mass scale $M_U$, and corresponding value of $\alpha^{-1}_U$  
(one exemplary case shown in the plot, $M_U=1.9 \times 10^{16}$ GeV and $\alpha^{-1}_U=41.4238$), 
we intend to estimate the proton life time $\tau_p$ and compare with the recent and proposed future experiments and also, if possible, derive uncertainties relevant for this result. The master formula for 
the gauge-induced $d=6$ proton decay in the chain $p \to e^+ \pi^0$ with the known heavy spectrum 
in this non-SUSY SO(10) model is
\begin{eqnarray}  
\Gamma\left( p\rightarrow \pi^0 e^+ \right) &=&\frac{\pi}{4}\, A^2_L\, \frac{|\overline{\alpha}_H|^2}{f^2_\pi} 
 \frac{m_p\, \alpha^2_U}{M^4_U}  \left(1 + \mathcal{F} + \mathcal{D} \right)^2 \mathcal{R}
\label{decay-width-proton}
\end{eqnarray}
where $A_L=1.25$ is the renormalization factor from the electroweak scale to the proton mass, $\mathcal{D}=0.81$, 
$\mathcal{F}=0.44$, $\overline{\alpha}_H=-0.011\, \, \mbox{GeV}^3$, and $f_\pi=139\, \, \mbox{MeV}$ are extracted as phenomenological parameters by chiral perturbation theory and lattice gauge theory. 
Also, $m_p=938.3\, \, \mbox{MeV}$ is the proton mass, and $\alpha_U \equiv \alpha_G$ is the gauge fine structure constant derived 
at the GUT scale. The renormalization factor $\mathcal{R}=\bigg[\left(A_{SR}^2+A_{SL}^2\right)
\left(1+ |{V_{ud}}|^2\right)^2\bigg]$ for $SO(10)$, the  $(1,1)$ element of $V_{CKM}$ is $V_{ud}=0.974=$ with $A_{SL}(A_{SR})$ being the short-distance 
renormalization factor in the left (right) sectors.

Redefining $\alpha_H = \overline{\alpha}_H \left(1 + \mathcal{F} + \mathcal{D} \right) = 0.012\, \, \mbox{GeV}^3$, and 
$\mathcal{A}_R\simeq \mathcal{A}_{L} \mathcal{A}_{SL} \simeq \mathcal{A}_{L} \mathcal{A}_{SL}$, the proton life time can 
be expressed as 
\begin{eqnarray}  
\tau_p = \Gamma^{-1}\left( p\rightarrow \pi^0 e^+ \right) &=&\frac{4}{\pi}\,\frac{f^2_\pi}{m_p}\frac{M^4_U}{\alpha^2_U} 
         \frac{1}{\alpha^2_H \mathcal{A}^2_{R}} \frac{1}{\mathcal{F}_q} \, ,
\label{lifetime-proton}
\end{eqnarray}
where $\mathcal{F}_q \simeq 7.6$

In order to estimate the proton life time, we should have knowledge about the short distance enhancement renormalization factors 
which are fully model dependent, a few of which are known while a few others have been already determined 
in the present model. For the particular choice of symmetry breaking considered in present non-SUSY SO(10) model and assuming no threshold corrections at or below the GUT scale, the short distance renormalization factors evaluated at one loop level are given as
\begin{eqnarray}
\mathcal{A}_{S}= \mathcal{A}^{2213D}_{S} \cdot \mathcal{A}^{2113}_{S} \cdot \mathcal{A}^{213}_{S}\, ,
\end{eqnarray}
where,  
\begin{eqnarray}
& &\mathcal{A}^{2213D}_{S}= \left(\frac{\alpha_{i} (M_P)}{\alpha_{i} (M_{U})} \right)
          ^{ -\frac{\gamma^{\prime \prime}_{i}}{2 \pmb a^{\prime \prime}_{i}}} 
          =\left(\frac{\alpha^{-1}_{i} (M_P)}{\alpha^{-1}_{i} (M_{U})} \right)
          ^{ \frac{\gamma^{\prime \prime}_{i}}{2 \pmb a^{\prime \prime}_{i}}}, \quad 
\mbox{i=2L, 2R, B-L, 3C} \, ; \nonumber    \\
& &\mathcal{A}^{2113}_{S}= \left(\frac{\alpha^{-1}_{i} (M^0_R)}{\alpha^{-1}_{i} (M_{P})} \right)
          ^{ \frac{\gamma^{\prime }_{i}}{2 \pmb a^{\prime}_{i}}}, \quad 
\mbox{i=2L, 1R, B-L, 3C} \, , 
\, ; \nonumber    \\
& &\mathcal{A}^{213}_{S}= \left(\frac{\alpha^{-1}_{i} (M_Z)}{\alpha^{-1}_{i} (M^0_{R})} \right)
          ^{ \frac{\gamma_{i}}{2 \pmb a_{i}}}, \quad 
\mbox{i=2L, Y, 3C} \, .
\end{eqnarray}
We have used the anomalous dimensions taken from \cite{anomalous:a,anomalous:b} and one-loop beta coefficients derived 
in our model. The estimated value of $\mathcal{A}_{R}=\mathcal{A}_{L} \cdot \mathcal{A}_{S}$ is $\mathcal{A}_{R} \simeq 2.24$. 
We have estimated the the proton life time to be $\tau_p = 5.75 \times 10^{35}$ yrs for the model under consideration with 
$M_U=1.9 \times 10^{16}$ GeV and $\alpha^{-1}=41.4238$. The predicted proton life time is out of reach for the current experiment 
Super-Kamiokande (2011) experiment giving bound on the proton life time for $p \to e^+ \pi^0$ channel is $\tau (p \to e^+ \pi^0) 
\big|_{SK, 2011} > 8.2 \times 10^{33}\, \mbox{yrs}$ \cite{Nishino:2012ipa}  while it can be accessible to future planned experiment such as 
$\tau (p \to e^+ \pi^0) \big|_{HK, 2025} > 9.0 \times 10^{34}\, \mbox{yrs}$ and $\tau (p \to e^+ \pi^0) \big|_{HK, 2040} > 2.0 
\times 10^{35}\, \mbox{yrs}$ \cite{babuetal}. 


\begin{figure}[h!]
\centering
\includegraphics[scale=0.6]{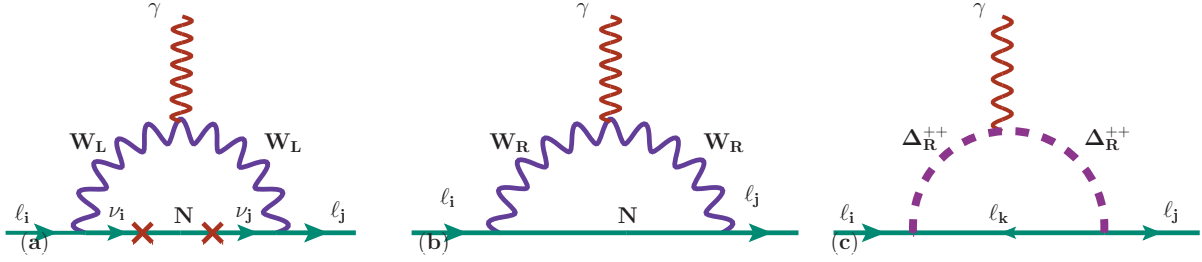}
\caption{One loop Feynman diagrams for lepton number violating decays $\ell_i \rightarrow \ell_j + \gamma (i \neq j)$. 
         Contribution from the $W_L$ exchanges involving mixing between left- and right-handed neutrinos is presented 
         in ({\bf a}) whereas contribution from the $W_R$ exchanges with heavy RH Majorana neutrinos is presented in 
         ({\bf b}). The dominant contribution to lepton flavor violation (LFV) decays via doubly charged RH Higgs triplet exchanges is presented 
         in ({\bf c}).}
\label{fig:LFV}
\end{figure}
\section{Lepton Flavor Violating Decays}
\label{LFV}
In the left-right model under consideration, there are different Feynman diagrams contributing to the underlying 
lepton-flavor violating interactions; (i) from $W_L$ exchanges with the mediation of light-heavy RH Majorana neutrinos 
shown in figure \,\ref{fig:LFV}(a), (ii) from $W_R$ exchanges with the mediation on heavy right-handed Majorana neutrinos shown 
in figure \,\ref{fig:LFV}(b), and (iii) from the doubly charged RH Higgs triplet ($\Delta^{++}_R$) exchanges as shown in 
figure \,\ref{fig:LFV}(c). The analytic expression for these contributions are given below
\begin{eqnarray}
& &\text{Br}\left(\mu \rightarrow e + \gamma \right)^{(a)}_{W_L} \simeq \frac{\alpha^3_W \sin^2\theta_W}{256 \pi^2} 
                 \frac{m^4_\mu}{M^4_{W_L}} \frac{m_\mu}{\Gamma_\mu} |\mathcal{G}^{\mu e}_{\gamma}|^2 \,, \nonumber \\
& &\text{Br}\left(\mu \rightarrow e + \gamma \right)^{(b)}_{W_R} \simeq \frac{3 \alpha_W}{32 \pi} \left(\frac{M_{W_L}}{M_{W_R}}\right)^{8} 
                 \left(\sin \theta_R\, \cos \theta_R \frac{M^2_2-M^2_1}{M^2_{W_L}}\right)^2 \,, \nonumber \\
& &\text{Br}\left(\mu \rightarrow e + \gamma \right)^{(c)}_{\Delta^{++}_R} \simeq \frac{2 \alpha_W\, M^4_{W_L}}{3 \pi g^4_{2R}} 
                  \left[ \frac{(f\, f^\dagger)_{12}}{M^2_{\Delta^{++}_R}}\right]^2\, , \nonumber 
\end{eqnarray}
where $\theta_W$ is the weak mixing angle, $\theta_R$ is the mixing angle between left and right handed neutrino sector, 
$\Gamma_\mu = 2.996 \times 10^{-19}\, \, \mbox{GeV}$, $\mathcal{G}^{\mu e}_{\gamma}$ contains left-right neutrino mixing 
plus the loop factor and $\alpha_W=g^2_{2L}/(4 \pi)$ is the fine structure constant for $SU(2)_L$ valid at $M_Z$ scale 
and is found to be $0.18389$. There have been several attempts to calculate the enhanced LFV 
signal in $\mu \to e \gamma$ process for example, in \cite{lfv-illakovac} and recently, it has been pointed out in refs.\,\cite{lfv-theory} 
that the LFV branching ratios can be significant if the heavy-light neutrino mixing is large. 
 
\begin{figure}[h!]
\centering
\includegraphics[scale=0.95]{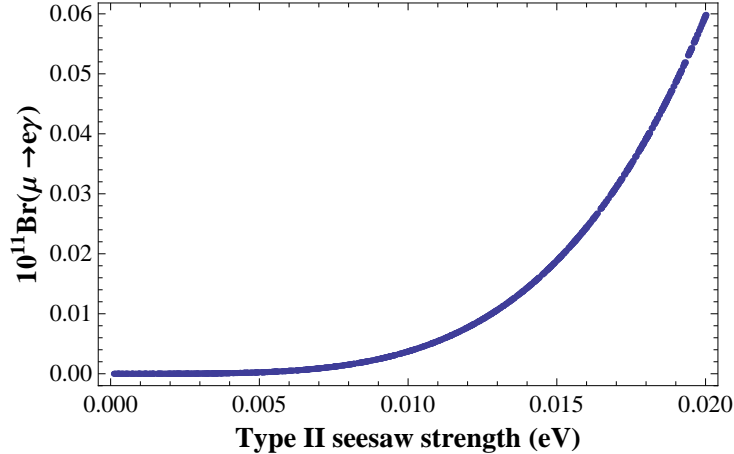}
\caption{Variation of the branching ration $\text{Br}\left(\mu \rightarrow e + \gamma \right)^{(c)}_{\Delta^{++}_R}$ arising 
        from the LFV decays via doubly charged RH Higgs triplet exchanges with the type II seesaw strength ($fv_L$).}
\label{fig:LFV-scatt}
\end{figure}

Assuming the left-right mixing to be small, one can neglect the contribution 
$\text{Br}\left(\mu \rightarrow e + \gamma \right)^{(a)}_{W_L}$ in comparison to other contributions. Also, in our model the $W_R$ gauge 
boson mass is found to be $\geq 10^{8}$ GeV making the $\text{Br}\left(\mu \rightarrow e + \gamma \right)^{(b)}_{W_R}$ 
contribution suppressed. The remaining dominant contribution due to TeV scale right-handed Higgs triplet contribution is 
\begin{eqnarray}
\label{eq3:LFV}
& &\text{Br}\left(\mu \rightarrow e + \gamma \right)^{(c)}_{\Delta^{++}_R} \simeq 
            \frac{2 \alpha_W\, M^4_{W_L}}{3 \pi g^4_{2R}} 
            \frac{1}{\left(\frac{\beta v^2 v_R}{M M_P} \right)^2}
             \left[ \frac{(m^{\small II}_\nu\, m^{{\small II}^\dagger}_{\nu})_{12}}{M^2_{\Delta^{++}_R}}\right]^2\,.
\end{eqnarray}
We have numerically estimated this contribution represented by a plot as shown in figure \,\ref{fig:LFV-scatt} 
where we have plotted $\text{Br}\left(\mu \rightarrow e + \gamma \right)^{(c)}_{\Delta^{++}_R}$ with the type II seesaw strength and using other allowed range of model parameters. From the plot, it can be seen that the numerical 
prediction for $\text{Br}\left(\mu \rightarrow e + \gamma \right)^{(c)}_{\Delta^{++}_R}$ in our model is same as the 
current MEG upper limit: $\text{Br}\left(\mu \rightarrow e + \gamma \right)^{(c)}_{\Delta^{++}_R}\bigg|_{\rm expt.} \leq 
5.7\times 10^{-13}$ \cite{MEG:expt,PRIME:expt} for type II seesaw strength $fv_L = 0.013$ eV. This is consistent with our model where the required type II seesaw strength is of the order of $0.001$ eV. 

\section{Conclusions}
\label{conclude}
We have studied a left-right symmetric gauge theory $SU(2)_L \times SU(2)_R \times U(1)_{B-L} \times SU(3)_C 
\times D (g_{2L}=g_{2R}) (\mathcal{G}_{2213D})$ which breaks down to the standard model gauge symmetry through 
two intermediate stages: first, the $SU(2)_R \times D$ breaks down to $U(1)_R$ at scale $M_P$ and  $U(1)_R 
\times U(1)_{B-L}$ breaks down to $U(1)_Y$ at a latter stage $M^0_R$. The motivation behind this set up 
is two-fold: $(i)$ to allow TeV scale intermediate $U(1)_R \times U(1)_{B-L}$ symmetry which can be accessed 
at experiments through $Z^{\prime}$, right handed neutrino and heavy Higgs searches, $(ii)$ to naturally 
allow type I seesaw dominance (which can give rise to TBM type $\mu-\tau$ symmetric neutrino mass matrix) 
while keeping type II seesaw term as sub-dominant but sizeable enough to give rise to the required deviation 
from TBM mixing in order to explain non-zero $\theta_{13}$. First we have performed a numerical analysis 
taking type I seesaw term as TBM type and type II seesaw term as a perturbation which breaks $\mu-\tau$ 
symmetry. We have done this exercise for both normal and inverted hierarchical neutrino mass spectra as 
well as two possible values of lightest neutrino mass (one being close to the maximum allowed by cosmological 
upper bound and one slightly lower). We have constrained the type II seesaw strength by demanding the required 
deviation from TBM to produce non-zero $\theta_{13}$. For dimensionless couplings to be of order one and 
$U(1)_R \times U(1)_{B-L}$ breaking scale of around $10$ TeV, the parity breaking scale has been restricted 
to be $10^9-10^{10}$ GeV.

We have also made an attempt to embed this model within $SO(10)$ GUT and check whether the above mentioned 
symmetry breaking steps can be naturally realized along with successful gauge coupling unification at a scale 
which lies close to the bound coming from proton lifetime constraint.  We have shown that in the framework 
of non-SUSY $SO(10)$ GUT invoking spontaneous D-parity breaking, one-loop RGE analysis of gauge couplings 
allow mass ranges $M^0_R=3-6$ TeV, $M_P=10^{9}-10^{11}$ GeV and $M_U=10^{14.5}-10^{16.5}$ GeV for several 
possible additional Higgs structures. We have also calculated the proton lifetime from the unification scale 
and find it to be within future experimental reach. At the end, we have made an estimate of branching ratio 
for the LFV decays of $\mu \rightarrow e + \gamma$ due to the presence of TeV scale doubly charged component 
of right handed triplet Higgs and found it to be lying close to the experimental limit.

\section{Acknowledgment} 
Two of the authors, Debasish Borah and Sudhanwa Patra would like to thank the organizers of the workshop 
entitled \textquotedblleft Majorana to LHC: Origin of neutrino Mass \textquotedblright at ICTP, Trieste, 
Italy during 2-5 October, 2013 where part of this work was completed. 



\begin{thebibliography}{50}
\bibitem{neutosc}
S.~Fukuda et al. (Super-Kamiokande),
{Phys. Rev. Lett.} {\bf 86}, 5656 (2001), hep-ex/0103033;
Q. R.~Ahmad et al. (SNO),
{Phys. Rev. Lett.} {\bf 89}, 011301 (2002), nucl-ex/0204008; 
{Phys. Rev. Lett.} {\bf 89}, 011302 (2002), nucl-ex/0204009;
J. N.~Bahcall and C.~Pena-Garay,
{New J. Phys.} {\bf 6}, 63 (2004), hep-ph/0404061; 
K. Abe et al. [T2K], {Phys. Rev. Lett.} {\bf 107}, 041801 (2011); 
P. Adamson et al. [MINOS], {Phys. Rev. Lett.} {\bf 107}, 181802 (2011);
Y. Abe et al. [DOUBLE-CHOOZ], {Phys. Rev. Lett.} {\bf 108}, 131801 (2012);
F. P. An et al. [DAYA-BAY], {Phys. Rev. Lett.} {\bf 108}, 171803 (2012);
J. K. Ahn et al. [RENO], {Phys. Rev. Lett.} {\bf 108}, 191802 (2012).
\bibitem{ti}
P. Minkowski, \textcolor{blue}{Phys.\, Lett.\, {\bf B\,67}, 421 (1977)}; 
M. Gell-Mann, P. Ramond, and R. Slansky (1980), print-80-0576 (CERN); 
T. Yanagida, Proceedings of the Workshop on the Baryon Number of the Universe and Unified Theories, Tsukuba, Japan, 13-14 Feb 1979; 
R.\,N. Mohapatra and G. Senjanovic, \href{http://dx.doi.org/10.1103/PhysRevLett.44.912}{\textcolor{blue}{Phys.\, Rev.\, Lett.\, {\bf 44}, 912 (1980)}}; 
J. Schechter and J.\,W.\,F. Valle, \href{http://dx.doi.org/10.1103/PhysRevD.22.2227}{\textcolor{blue}{Phys.\,Rev.\, {\bf D\,22}, 2227 (1980)}}.

\bibitem{tii} 
R.\,N. Mohapatra and G. Senjanovic, \textcolor{blue}{Phys.\,Rev.\, {\bf D\,23}, 165 (1981)}; 
G. Lazarides, Q. Shafi and C Wetterich, \textcolor{blue}{Nucl.\,Phys.\,{\bf B\,181}, 287 (1981)}; 
C. Wetterich, \textcolor{blue}{Nucl.\,Phys.\,{\bf B\,187}, 343 (1981)}; 
S. Antusch and S.\,F.\, King, \textcolor{blue}{Phys.\,Lett.\,{\bf B\,597}, 199 (2004)}.

\bibitem{LR}
J. C. Pati and A. Salam, Phys. Rev. {\bf D10}, 275 (1974); 
R. N. Mohapatra and J. C. Pati, Phys. Rev. {\bf D11}, 2558 (1975); 
G. Senjanovic and R. N. Mohapatra, Phys. Rev. {\bf D12}, 1502 (1975); 
R. N. Mohapatra and R. E. Marshak, Phys. Rev. Lett. {\bf 44}, 1316 (1980); 
N. G. Deshpande, J. F. Gunion, B. Kayser, and F. I. Olness, Phys. Rev. {\bf D44}, 837 (1991).

\bibitem{tello}
V. Tello, M. Nemevsek, F. Nesti, G. Senjanovic and F. Vissani, Phys. Rev. Lett. {\bf 106}, 151801 (2011).
\bibitem{patra-jhep}
R. L. Awasthi, M. K. Parida and S. Patra, JHEP {\bf 1308}, 122 (2013).

\bibitem{Dparity11}
D. Chang, R.\,N. Mohapatra, and M.\,K. Parida, \textcolor{blue}{Phys.\,Rev.\,Lett.\,{\bf 52}, 1072 (1984)}; 
D. Chang, R.\,N. Mohapatra, and M.\,K. Parida, \textcolor{blue}{Phys.\,Rev.\,{\bf D\,30}, 1052 (1984)}.

\bibitem{Dparity22}
D. Chang, R.\,N. Mohapatra, J. Gipson, R.\,E. Marshak, and M.\,K. Parida, \textcolor{blue}{Phys. Rev. {\bf D\, 81}, 1718 (1985)}.

\bibitem{patra-debi}
D. Borah, S. Patra, and U. Sarkar, 
\href{http://dx.doi.org/10.1103/PhysRevD.83.035007}{\textcolor{blue}{Phys.\,Rev.\,{\bf D\,83}, 035007 (2011)}}, 
\textcolor{blue}{arXiv:1006.2245 [hep-ph]}.   



\bibitem{utpal-sahu}
N. Sahu and U. Sarkar, 
        \href{http://dx.doi.org/10.1103/PhysRevD.74.093002}{\textcolor{blue}{Phys.\,Rev.\,{\bf D\,74}, 093002 (2006)}}, 
        \textcolor{blue}{arXiv:0605007 [hep-ph]}.

\bibitem{utpal-desai}
K. Bhattacharya, C. R. Das, B. R. Desai, G. Rajasekaran and U. Sarkar, 
        \href{http://dx.doi.org/10.1103/PhysRevD.74.015003}{\textcolor{blue}{Phys.\,Rev.\,{\bf D\,74}, 015003 (2006)}}, 
        \textcolor{blue}{arXiv:0601170 [hep-ph]}.
        
\bibitem{parida1312}  
B. P. Nayak and M. K. Parida, \textcolor{blue}{arXiv:1312.3185 [hep-ph]}

\bibitem{schwetz12} 
M.\,C. Gonzalez-Garcia, M. Maltoni, J. Salvado and T. Schwetz, 
\textcolor{blue}{JHEP {\bf 1212}, 123 (2012)}, \textcolor{blue}{arXiv:1209.3023 [hep-ph]}.
\bibitem{Harrison} 
P.\,F. Harrison, D.\,H. Perkins and W.\,G. Scott, \textcolor{blue}{Phys.\,Lett.\,{\bf B\,530}, 167 (2002)}; 
P.\,F. Harrison and W.\,G. Scott, \textcolor{blue}{Phys.\,Lett.\, {\bf B\,535}, 163 (2002)}; 
Z.\,Z. Xing, \textcolor{blue}{Phys.\,Lett.\,{\bf B\,533}, 85 (2002)}; 
P.\,F. Harrison and W.\,G. Scott, \textcolor{blue}{Phys.\,Lett.\,{\bf B\,547}, 219 (2002)}; 
P.\,F. Harrison and W.\,G. Scott, \textcolor{blue}{Phys.\,Lett.\,{\bf B\,557}, 76 (2003)}; 
P.\,F. Harrison and W.\,G. Scott, \textcolor{blue}{Phys.\,Lett.\,{\bf B\,594}, 324 (2004)}.


\bibitem{nzt13} 
Y. Shimizu, M. Tanimoto and A. Watanabe, 
        \textcolor{blue}{Prog.\,Theor.\,Phys.\,{\bf 126}, 81 (2011)}; 
S.\,F. King and C. Luhn, 
         \textcolor{blue}{JHEP {\bf 1109}, 042 (2011)}; 
S. Antusch, S.\,F. King, C. Luhn and M. Spinrath, 
         \textcolor{blue}{Nucl.\,Phys.\,{\bf B\,856}, 328 (2012)}; 
S.\,F. King and C. Luhn, 
         \textcolor{blue}{JHEP {\bf 1203}, 036 (2012)}; 
S. Gupta, A.\,S. Joshipura and K.\,M. Patel, 
         \textcolor{blue}{Phys.\,Rev.\,{\bf D\,85}, 031903 (2012)}; 
S-F. Ge, D.\,A. Dicus and W.\,W. Repko, 
         \textcolor{blue}{Phys.\,Rev.\,Lett.\,{\bf 108}, 041801 (2012)}; 
S-F. Ge, D.\,A. Dicus and W.\,W. Repko, 
         \textcolor{blue}{Phys.\,Lett.\,{\bf B\,702}, 220 (2011)}; 
M-C. Chen, J. Huang, J-M. O'Bryan, A.\,M. Wijangco and F. Yu, 
         \textcolor{blue}{JHEP {\bf 1302}, 021 (2012)}.
         
\bibitem{nzt13GA} 
G. Altarelli, F. Feruglio, L. Merlo and E. Stamou, 
         \textcolor{blue}{JHEP {\bf 1208}, 021 (2012)}.
         
\bibitem{devtbmt2} 
W. Rodejohann, 
         \textcolor{blue}{Phys.\,Rev.\,{\bf D\,70}, 073010 (2004)}; 
M. Lindner and W. Rodejohann, 
          \textcolor{blue}{JHEP {\bf 0705}, 089 (2007)}; 
D.\,A. Sierra and I.\,de\,M. Varzielas and E. Houet, 
          \textcolor{blue}{Phys.\,Rev.\, {\bf D\,87}, 093009 (2013)}.

\bibitem{dborah7-13} 
D. Borah, \textcolor{blue}{Nucl.\,Phys.\,{\bf B\,876}, 575 (2013)}.



\bibitem{Planck13} 
P.\,A.\,R. Ade {\em et al.}, {\bf (Planck Collaboration)}, \textcolor{blue}{arXiv:1303.5076 [hep-ph]}.

\bibitem{PDG}
C. Amsler et al. (Particle Data Group), Phys. Lett. B667, 1 (2008).

\bibitem{borahdwall}
D. Borah, \textcolor{blue}{Phys.\, Rev.\, {\bf D86}, 096003 (2012)}.

\bibitem{anomalous:a}
A.\,J. Buras, J. Ellis, M.\,K. Gaillard and D.\,V. Nanopoulos,  
\textcolor{blue}{Nucl.\,Phys.\,{\bf B\,135}, 66 (1978)}; 
J. Ellis, D.\,V. Nanopoulos and S. Rudaz, 
\textcolor{blue}{Nucl.\,Phys.\,{\bf B\,202}, 43 (1982)}; 
T. Goldman and D.\,A. Ross, 
\textcolor{blue}{Nucl.\,Phys.\,{\bf B\,171}, 273 (1980)}; 
J. Hisano, H. Murayama and T. Yanagida, 
\textcolor{blue}{Nucl.\,Phys.\,{\bf B\,402}, 46 (1993)}; 
L.\,E. Ibanez and C. Munoz, 
\textcolor{blue}{Nucl.\,Phys.\,{\bf B\,245}, 425 (1984)}.
\bibitem{anomalous:b}
J. Hisano, D. Kobayashi and N. Nagata, 
\href{http://dx.doi.org/10.1016/j.physletb.2012.08.037}{\textcolor{blue}{Phys.\,Lett.\, {\bf B\,716}, 406 (2012)}}. 
\textcolor{black}{arXiv:1204.6274 [hep-ph]}.

S.\, Bertolini, L.-Di Luzio and M. Malinsky
%
%




\bibitem{Nishino:2012ipa}
{\bf Super-Kamiokande Collaboration}, H.~ Nishino {\em et al.}, 
    \href{http://dx.doi.org/10.1103/PhysRevD.85.112001}{\textcolor{blue}{Phys.\,Rev. {\bf D\,85} (2012) 112001}}.
      \href{http://arxiv.org/abs/1203.4030}{\textcolor{black}{arXiv:1203.4030 [hep-ph]}}.

\bibitem{babuetal} 
K. S. Babu {\em et al.}, 
      \href{http://arxiv.org/abs/1205.2671}{\textcolor{blue}{arXiv:1205.2671 [hep-ex]}}.
\bibitem{lfv-illakovac}
W.\,J. Marciano and A.\,I. Sanda, 
        \textcolor{black}{Phys.\,Lett.\,{\bf B\,67}, 303 (1977)}; 
T.\,P. Cheng and L.\,-F. Li,
        \textcolor{black}{Phys.\,Rev.\,Lett.\,{\bf 45}, 1908 (1980)}; 
P. Langacker and D. London, 
        \textcolor{black}{Phys.\,Rev.\,{\bf D\,38}, 907 (1988)}. 
A.~Ilakovac, A.~Pilaftsis,  \textcolor{blue}{Nucl.\,Phys.\,{\bf B\,437}, 491 (1995)}, 
        \textcolor{black}{arXiv:9403398 [hep-ph]}; 

\bibitem{lfv-theory}
R. Alonso, M. Dhen, M.\,B. Gavela and T. Hambye, 
           \textcolor{blue}{JHEP\,{\bf 1301}, 118 (2013)}; 
Chang-Hun Lee, P.\,S.Bhupal Dev, and R.\,N. Mohapatra, 
           \textcolor{blue}{Phys.\,Rev.\,{\bf D\,88}, 093010 (2013)}.
           \textcolor{black}{arXiv:1309.0774 [hep-ph]}.
\bibitem{PRIME:expt}
J.\, Adam {\em et al}, {\bf MEG Collaboration}, 
         \textcolor{blue}{arXiv:1303.0754 [hep-ex]}; 
A.\,M. Baldini {\em et al}, {\bf MEG Collaboration}, 
         \textcolor{blue}{arXiv:1301.7225 [physics.ins-det]}. 
\bibitem{MEG:expt} 
Y. Kuno {\em et al}, {\bf PRISM, PRIME Collaboration}, 
         \textcolor{blue}{Nucl.\,Phys.\,Proc.\,Suppl.\, {\bf 149}, 376 (2005)}.
        
        
\end{thebibliography}
\end{document}